\documentclass[aps,twocolumn,english,superscriptaddress,citeautoscript,preprintnumbers,amsmath,amssymb,floatfix,footinbib]{revtex4-1}
\usepackage[breaklinks=true,colorlinks,citecolor=blue,linkcolor=blue,urlcolor=blue]{hyperref}
\usepackage{amsmath}
\usepackage{graphicx}
\usepackage{dcolumn}
\usepackage{float}
\usepackage[utf8]{inputenc}
\usepackage{color}
\usepackage[super]{nth}
\usepackage{soul}

\begin{document}
	
\title{Electronic structure and physical properties of EuAuAs single crystal}
\author{S. Malick}
\affiliation{Department of Physics, Indian Institute of Technology, Kanpur 208016, India}
\author{J. Singh}
\affiliation{Department of Physics, Indian Institute of Technology Hyderabad, Kandi, Medak 502 285, Telangana, India}
\author{ A. Laha}
\affiliation{Department of Physics, Indian Institute of Technology, Kanpur 208016, India}
\author{V. Kanchana}
\email{kanchana@iith.ac.in}
\affiliation{Department of Physics, Indian Institute of Technology Hyderabad, Kandi, Medak 502 285, Telangana, India}

\author{Z. Hossain}
\email{zakir@iitk.ac.in}
\affiliation{Department of Physics, Indian Institute of Technology, Kanpur 208016, India}
\affiliation{Institute of Low Temperature and Structure Research,
	Polish Academy of Sciences, ulica Okolna 2, 50-422 Wroclaw, Poland}
\author{D. Kaczorowski}
\email{d.kaczorowski@intibs.pl}
\affiliation{Institute of Low Temperature and Structure Research,
	Polish Academy of Sciences, ul. Okolna 2, 50-422 Wroclaw, Poland}

\begin{abstract}
	
High-quality single crystals of EuAuAs were studied by means of powder x-ray diffraction, magnetization, magnetic susceptibility, heat capacity, electrical resistivity and magnetoresistance measurements. The compound crystallizes with a hexagonal structure of the ZrSiBe type (space group $P6_3/mmc$). It orders antiferromagnetically below 6 K due to the magnetic moments of divalent Eu ions. The electrical resistivity exhibits metallic behavior down to 40 K, followed by a sharp increase at low temperatures. The magnetotransport isotherms show a distinct metamagnetic-like transition in concert with the magnetization data. The antiferromagnetic ground state in \mbox{EuAuAs} was corroborated in the \textit{ab initio} electronic band structure calculations. Most remarkably, the calculations revealed the presence of nodal line without spin-orbit coupling and Dirac point with inclusion of spin-orbit coupling. The \textit{Z}$_2$ invariants under the effective time reversal and inversion symmetries make this system nontrivial topological material. Our findings, combined with experimental analysis, makes EuAuAs a plausible candidate for an antiferromagnetic topological nodal-line semimetal.

\end{abstract}
	
	\maketitle
	
Among rare-earth intermetallics, Eu-based compounds have drawn special attention as they exhibit a variety of magnetic ground states with large Eu$^{2+}$ moment. A well-known example of Eu-based compound is EuFe$_2$As$_2$, which has been studied enormously because of the coexistence of superconductivity (SC) and magnetism. SC arises in this compound due to the suppression of spin density wave transition by applying hydrostatic or chemical pressure, which results in a fascinating phase diagram connecting superconductivity and magnetism as a function of external pressure or doping \cite{EuFe2As2_PRB_2008, EuFe2As2_PRB_2008_Possible_SC, EuFe2As2_PRB_2019_Pressure_Neutron, EuFe2As2_JPCM_K_Doped_2011, EuFe2As2_PRB_Ni_Doped_2012, EuFe2As2_SST_Ir_doped_2014}. Similarly, Eu-based materials having EuT$_2$X$_2$ \cite{EuT2X2} and EuTX (T = transition metals, X = post-transition metals and metalloid) \cite{Eu_ternary_111_Pottgen,Divalent_Eu,Eu_111,EuPdAs,EuZnSn,EuCuAs} chemical composition have also been well studied because of their complex magnetic structures. Recently, there has been a new urge to search for novel magnetic topological materials, as they offer a few extraordinary features such as chiral magnetic anomaly, anomalous Hall effect, and anomalous Nernst effect \cite{Weyl_review,PrAlGe,GdPtBi,Co3SnS2,Co2MnGa,Co2MnGa_2019,Fe3GeTe2,Mn3Sn_Nature_2015}. The magnetism breaks the time-reversal symmetry (TRS) of an electronic structure and such a nontrivial system with broken TRS may induce a large anomalous Hall effect, influenced by large Berry curvature \cite{Co3SnS2,Fe3GeTe2,RAlGe}. Furthermore, noncoplanar magnetic spin texture in the topologically nontrivial state generates the topological Hall effect \cite{CeAlGe,Mn3Sn}. Despite intensive research in the past few years, the external magnetic field tunable real-space spin-texture in Weyl semimetals (WSMs) or topological insulators (TIs) remains largely unexplored.

Recently, ternary pnictides with hexagonal structures have drawn significant attention as this group of compounds offers various topological states. For instance, a first-principle calculation suggested that BaAgAs is a Dirac semimetal (DSM) with pair of Dirac points lying on the $C_3$ rotation axis. Further, potassium doping in BaAgAs can transfer the system into a triple-point semimetal (TPSM) state \cite{BaAgAs}. Likewise, a controlled amount of Cu doping in the Ag site of SrAgAs may induce discrete topological states, e.g.,  DSM, TPSM, WSM, and TI \cite{SrAgAs}. Antiferromagnetic DSM compound EuAgAs exhibits chiral anomaly induced negative longitudinal magnetoresistance (MR) and large topological Hall effect \cite{EuAgAs}. In contrast, the magnetotransport in the iso-structural compound EuCuAs does not show any topological features \cite{EuCuAs}. Another nonmagnetic compound CaAuAs from the same family was experimentally verified as DSM \cite{CaAuAs_ARPES}. Moreover, nonmagnetic topological nodal-line semimetals CaCdX (X = Ge and Sn) exhibit several interesting magnetotransport properties such as large nonsaturating MR, magnetic field induced metal semiconductorlike crossover and a plateau in resistivity at low temperatures \cite{CaTX,CaCdSn}. Interestingly, strong electron correlations have been found in the nodal-line semimetals, YbCdX (X= Ge and Sn) \cite{YbCdGe,YbCdSn}.

The present work was motivated by the findings reported for CaAuAs. We replaced Ca with Eu while maintaining the crystal symmetry in order to investigate the interplay between magnetism and topology. Here, we report the results of our magnetic, magnetotransport, and heat capacity measurements on EuAuAs single crystals. The magnetic data suggest that EuAuAs orders antiferromagnetically below 6 K. A pronounced anomaly in the specific heat near 6 K corroborates the bulk magnetic ordering. The electrical resistivity decreases monotonically with decreasing temperature down to 40 K, then rises sharply and eventually forms a plateau at low temperatures.  Our magnetotransport study revealed large negative MR. The experimental characterization of EuAuAs was accompanied by bulk electronic band structure calculations and surface state analysis, which revealed the presence of a nodal line. Based on these results we can conclude that \mbox{EuAuAs} is a novel antiferromagnetic nodal-line semimetal.
\begin{figure}
	\includegraphics[width=8.4cm, keepaspectratio]{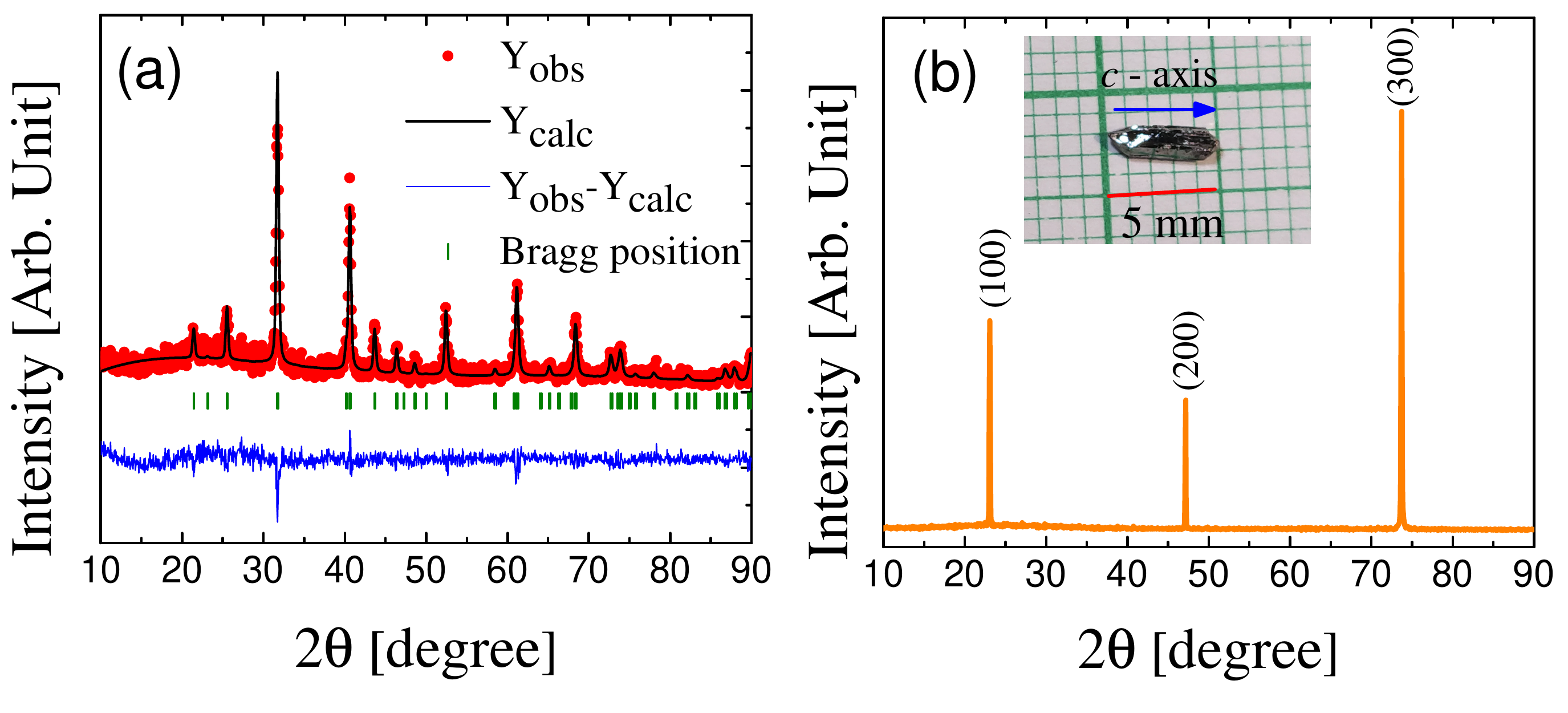}
	\caption{\label{Fig1}(a) Powder XRD pattern of EuAuAs recorded at room temperature. The observed and calculated diffractograms are indicated by red points and solid black line, respectively. Solid blue line represents the difference between the experimentally observed and calculated intensities. Green tick marks are Bragg peak positions. (b) Single-crystal XRD pattern. Inset shows a photograph of a typical EuAuAs single crystal.}
\end{figure}

\section{Experimental Details and Methods}

The single crystals of EuAuAs were grown using bismuth flux. Individual elements Eu (99.9\%, Alfa Aesar), Au (99.999\%, Alfa Aesar), As (99.999\%, Alfa Aesar), and Bi (99.99 \%, Alfa Aesar) were taken in 1:1:1:10 molar ratio. All the elements were put into an alumina crucible that was sealed in a quartz tube with partial argon pressure. The ampoule was heated to 1050 $^\circ$C for 20 h, and then cooled down to 600 $^\circ$C at a rate 3 $^\circ$C/h. At this temperature, the crystals were separated from the flux using a centrifuge. The crystals were of typical size 3 $\times$ 1 $\times$ 1 mm$^3$. The crystal structure was investigated by x-ray diffraction (XRD) using a PANalytical X'Pert PRO diffractometer with Cu K$_{\alpha1}$ radiation. Their chemical composition was examined by energy dispersion x-ray spectroscopy (EDS) employing a JEOL JSM-6010LA electron microscope. Electrical resistivity and magnetoresistance measurements were performed by conventional four-probe technique using a Quantum Design Physical Property Measurement System (PPMS). Heat capacity measurements were conducted in the same PPMS platform. Magnetic measurements were carried out using a Quantum Design Magnetic Property Measurement System.

The room temperature powder and single-crystal XRD patterns are shown in Figs. \ref{Fig1}(a) and \ref{Fig1}(b), respectively. The XRD data confirm that EuAuAs crystallizes in ZrSiBe-type hexagonal structure with the space group $P6_3/mmc$ (no. 194). The lattice parameters obtained from the Rietveld refinement of the powder XRD data are $a$ = $b$ = 4.441 \AA~ and $c$ = 8.289 \AA, in good agreement with the previous report \cite{Eu_ternary_111_Pottgen}. The EDS data confirm the expected equiatomic stoichiometry of the crystals investigated.  

For the electronic band structure calculations, a geometry optimization was performed using a pseudo potential method implemented in the {\footnotesize VASP} package within the framework of density functional theory \cite{Kresse_B47, Kresse_6, Kresse_B54, Kresse_B59}. The PBE-GGA pseudo potential was used for the exchange correlation functional \cite{Perdew}. The strong correlation effects of Eu-\textit{f} states were treated by applying an effective Hubbard \textit{U} parameter (GGA+\textit{U}) as \textit{U} = 7 eV \cite{WanU7, KuneU7}. The plane wave energy cutoff was set to 500 eV for all calculations. The energy convergence criterion was chosen to be 10$^{-6}$ eV. According to the Monkhorst-Pack scheme, 8$\times$8$\times$4 \textit{k} mesh was used for the geometry optimization calculations \cite{Monkhorst}. For the surface state calculations, tight-binding Hamiltonian was obtained based on maximally localized Wannier functions using the Wannier90 package \cite{G_Pizzi}. Based on the tight-binding model, the iterative Green’s function method, which is implemented in the WannierTools package, was used to investigate topological properties of the compound \cite{Q_Wu, MPL}.

\section{Results and Discussion}
\subsection{Magnetic properties}

\begin{figure*}[htb]
	\centering
	\begin{tabular}{@{}cccc@{}}
		\includegraphics[width=0.335\textwidth]{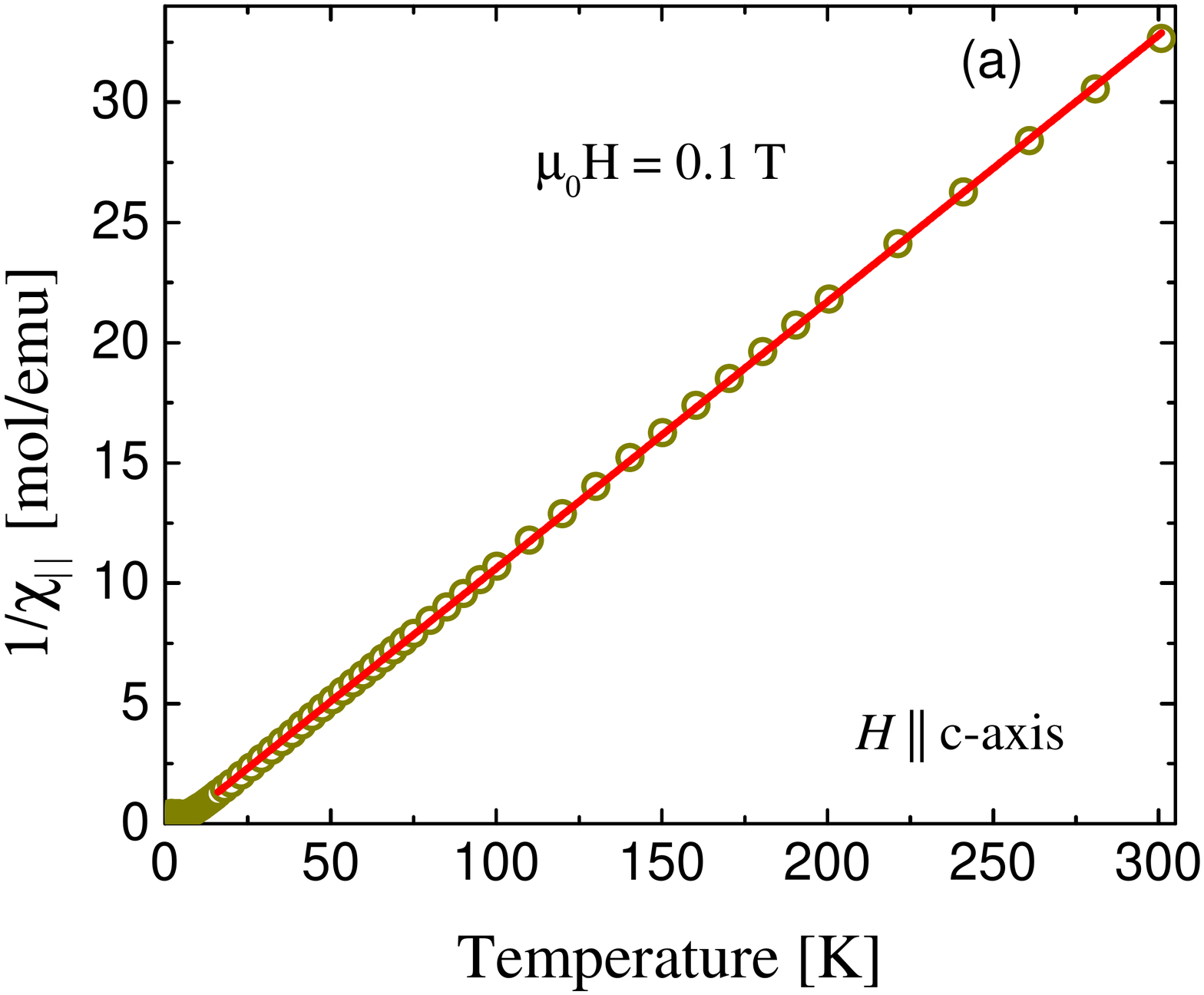} &
		\includegraphics[width=0.335\textwidth]{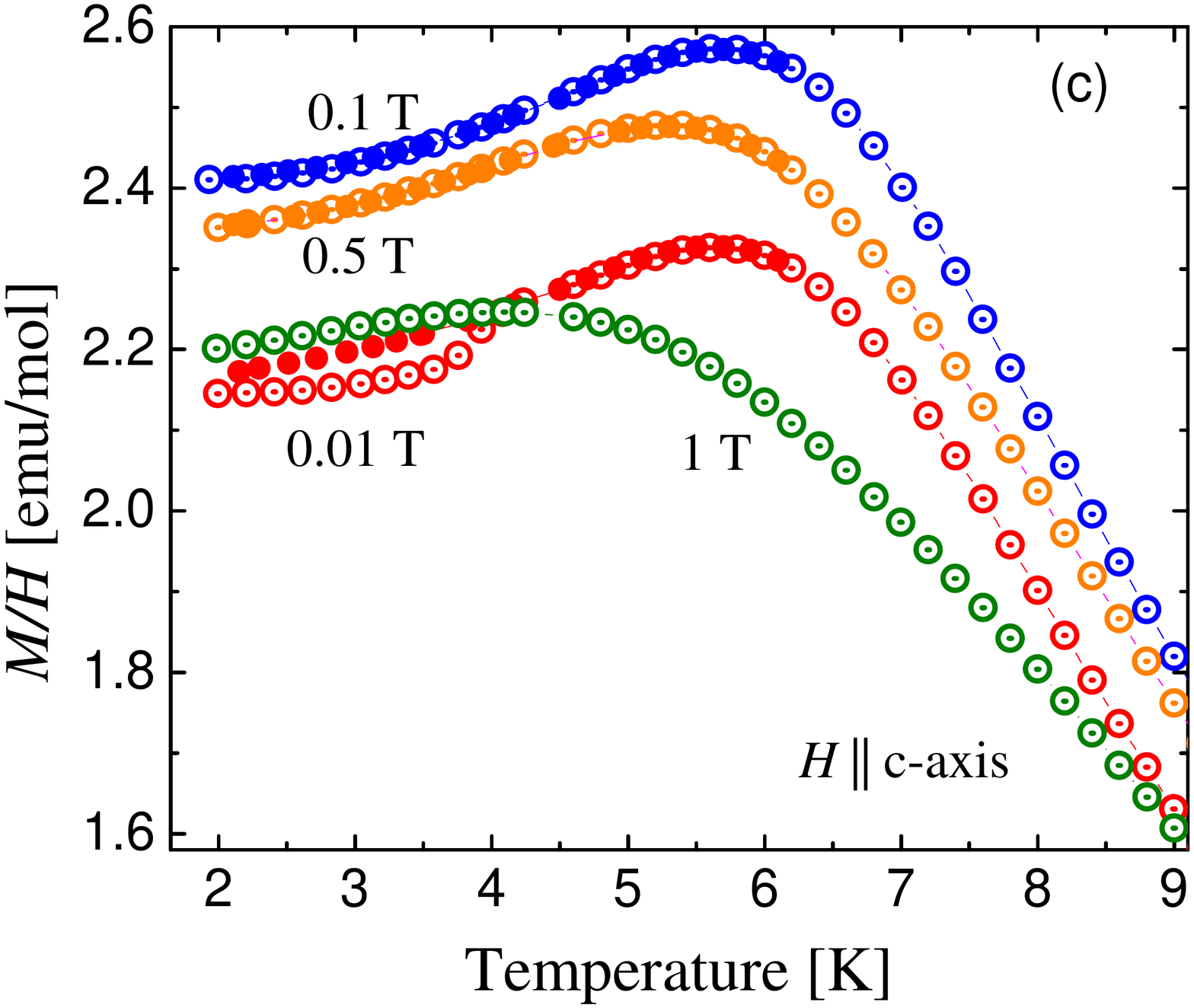} &
		\includegraphics[width=0.335\textwidth]{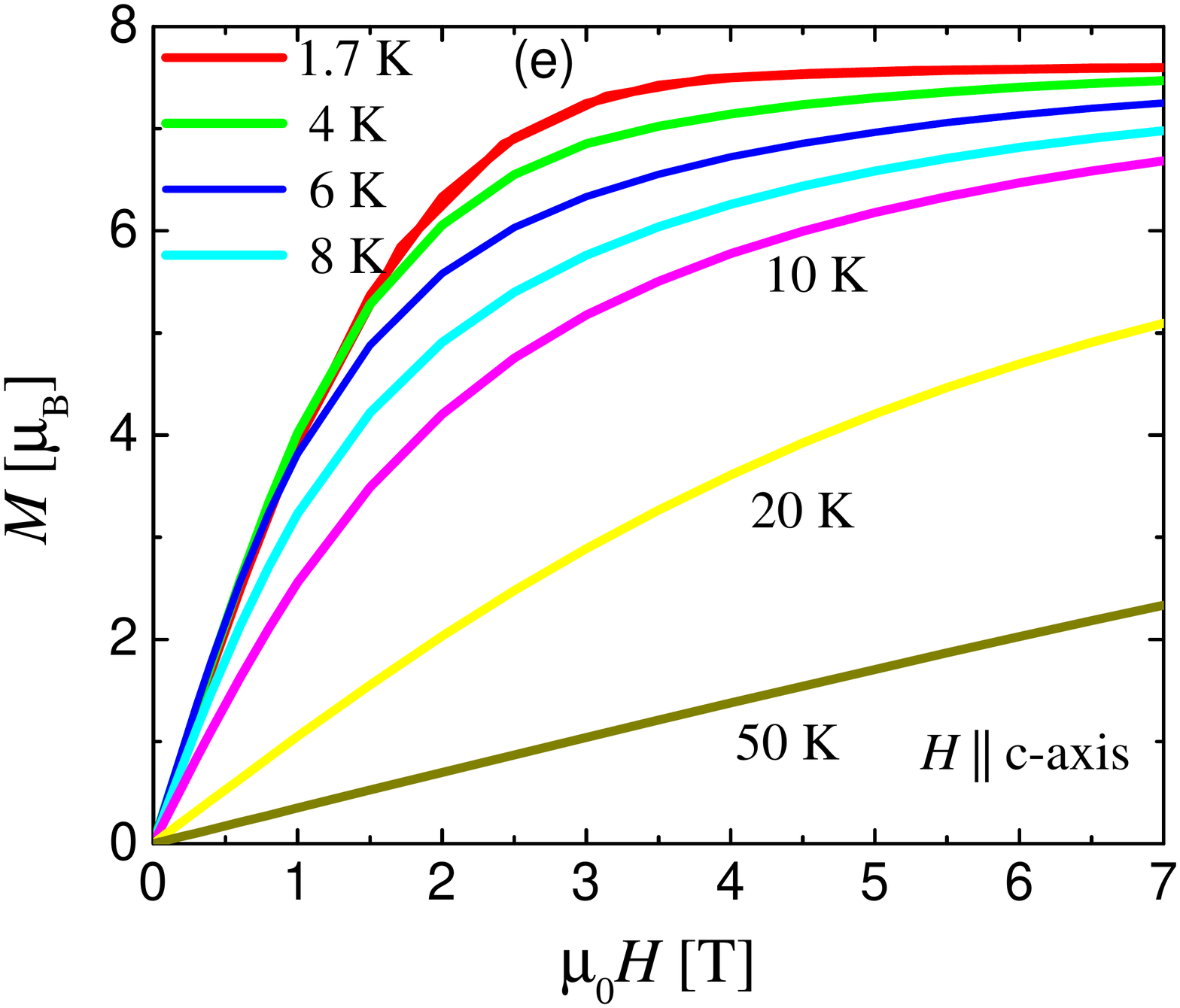} \\
		\includegraphics[width=0.335\textwidth]{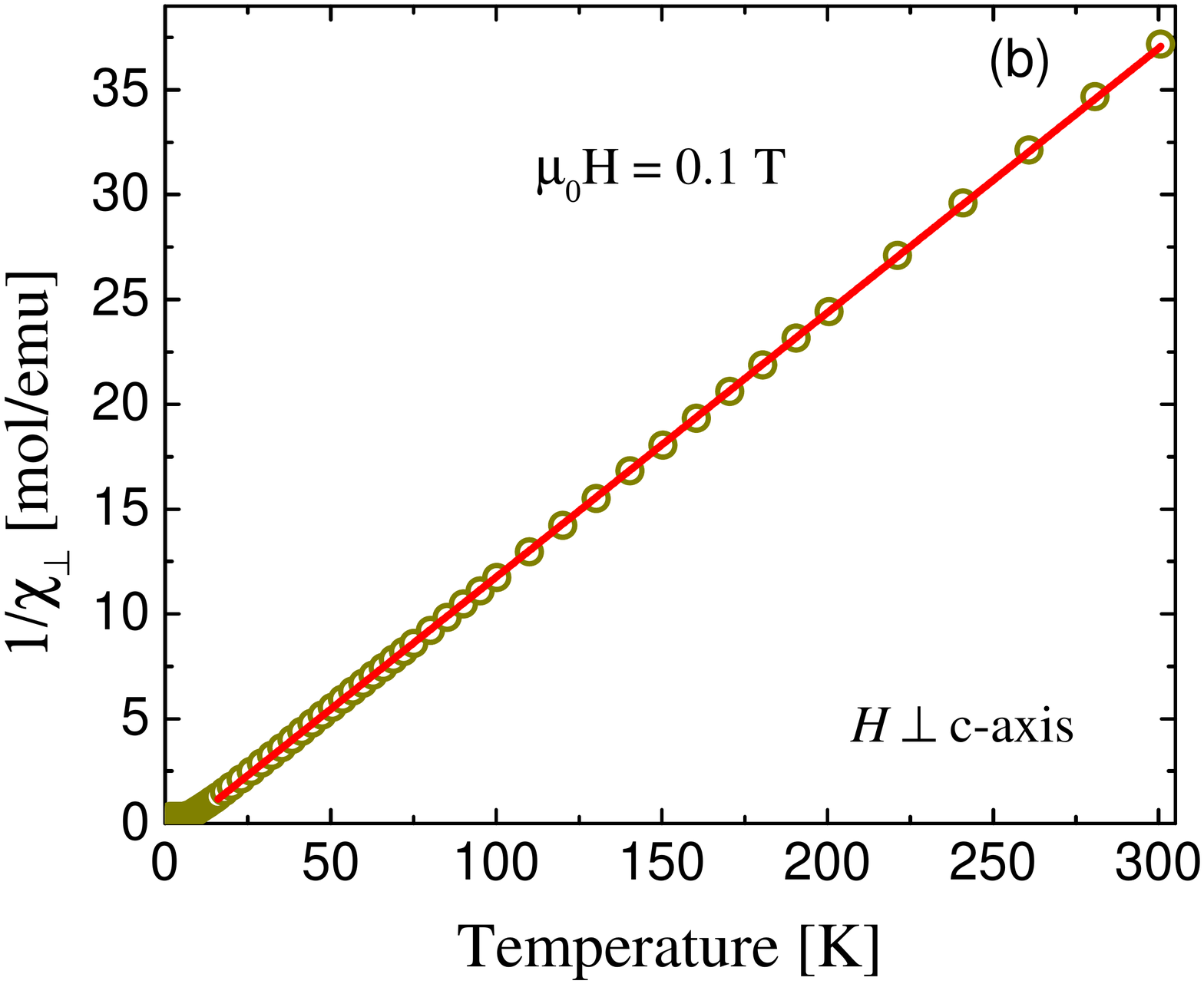} &
		\includegraphics[width=0.335\textwidth]{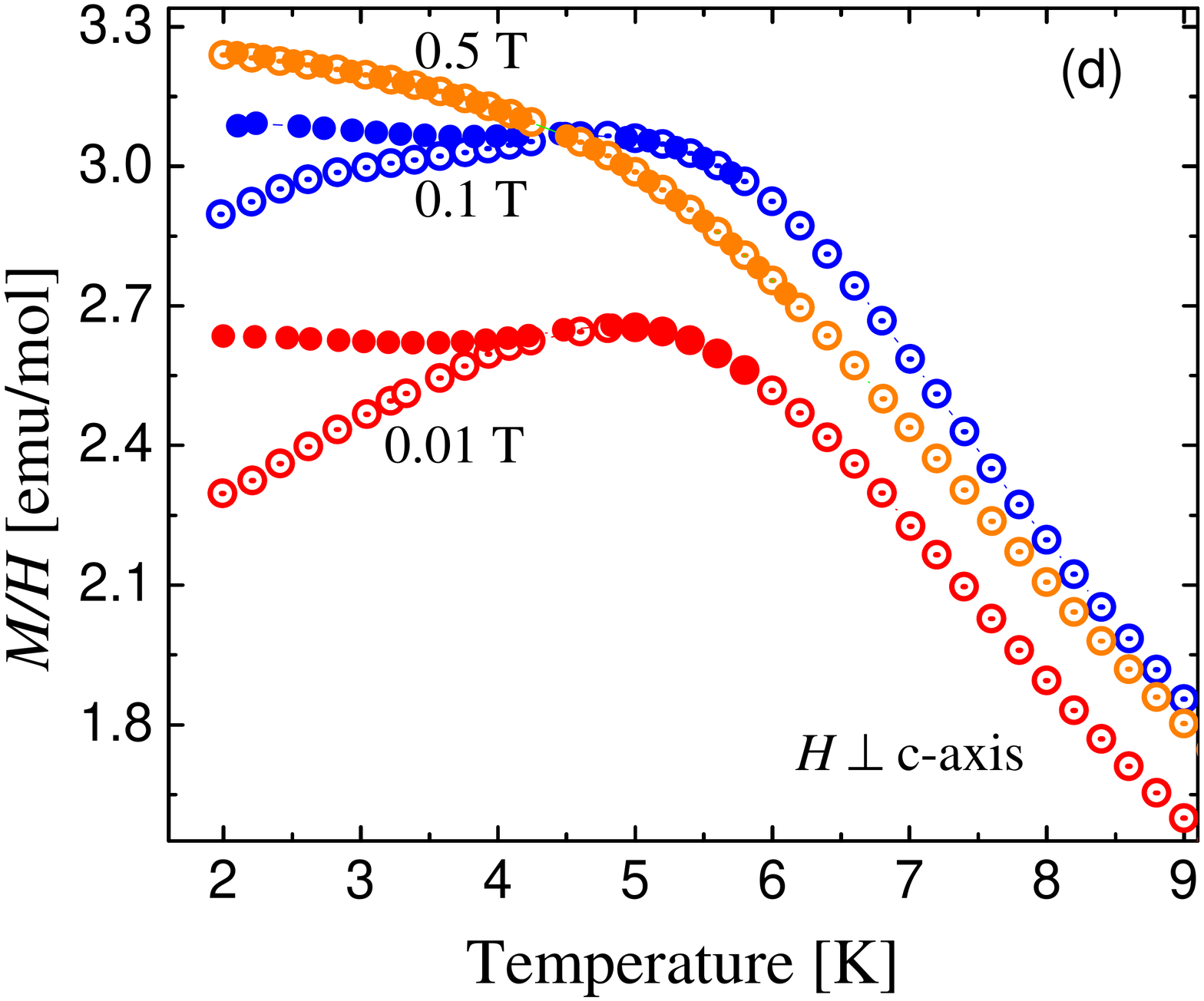}&
		\includegraphics[width=0.335\textwidth]{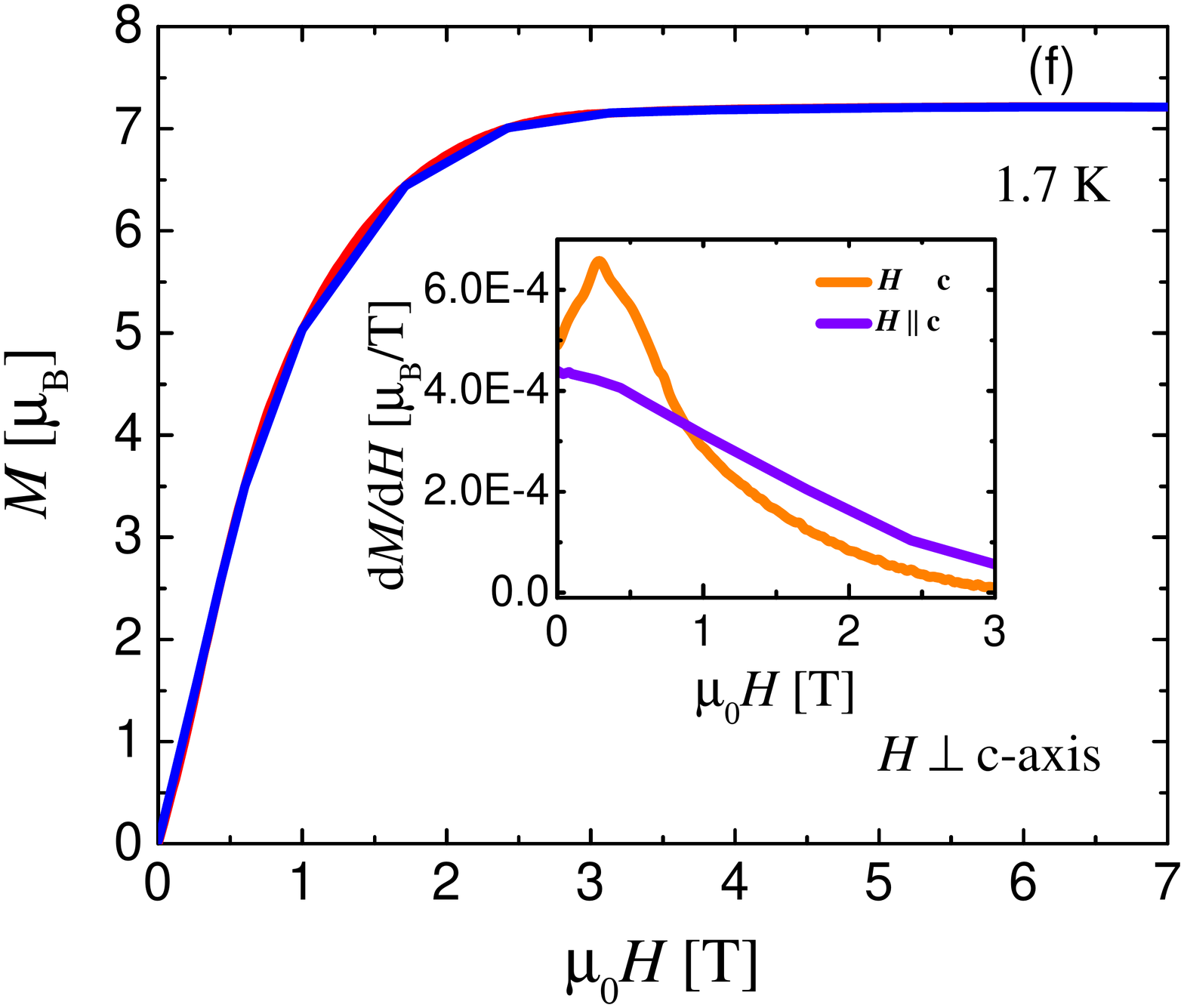}
		
	\end{tabular}
	\caption{\label{Fig2} Magnetic properties of single-crystalline EuAuAs:  Temperature-dependent inverse magnetic susceptibility measured in a magnetic field of 0.1 T along \textit{c} axis and perpendicular to the \textit{c} axis are shown in panels (a) and (b), respectively. Panels (c) and (d) present the variations of $M/H$ with respect to temperature, measured in FC (filled circles) and ZFC (open circles) regimes in different magnetic fields applied along the \textit{c} axis and perpendicular to the \textit{c} axis, respectively. (e) Field-dependent magnetization measured at different temperatures along the \textit{c} axis. (f) Magnetization isotherm measured at 1.7 K for field perpendicular to \textit{c} axis with increasing (red line) and decreasing (blue line) magnetic field. Inset represents the temperature derivative of the magnetization data in weak fields region for $H\perp c$ and $H\parallel c$.}
\end{figure*}

Zero field cooled (ZFC) inverse magnetic susceptibility $\chi_{\parallel}(T)$ and $\chi_{\perp}(T)$ measured on a single crystal of EuAuAs in a magnetic field $\mu_0 H$ of 0.1 T for $H$$\parallel c$ and $H$$\perp$\textit{c} axis, respectively, in the temperature range 1.7-300 K is shown in Figs. \ref{Fig2}(a) and \ref{Fig2}(b), respectively. The inverse susceptibility data can be described by the Curie-Weiss law,  $\chi(T) = C/(T-\Theta_P)$; where $C$ and  $\Theta_P$ are the Curie constant and paramagnetic Curie temperature, respectively. The obtained values of effective moments are 8.53$\mu_B$ ($H$$\parallel c$) and 8.00$\mu_B$ ($H$$\perp c$), i.e. close to the theoretical one, $g\sqrt{S(S+1)} \mu_B=7.94\mu_B$ ($S$= 7/2 and $g$ = 2) for a free Eu$^{2+}$ ion. The obtained values of $\Theta_P$ are 4.1 K ($H$$\parallel c$) and 6.7 K ($H$$\perp c$). Their positive sign indicates predominance of ferromagnetic exchange interaction in EuAuAs. The magnetization $M$ measured in ZFC and field cooled (FC) regimes in different magnetic fields applied along and perpendicular to the \textit{c} axis are plotted as $M/H$ vs. $T$ in Figs. \ref{Fig2}(c) and \ref{Fig2}(d), respectively. A clear peak around 6 K in both data indicates an antiferromagnetic (AFM) phase transition. With increasing applied field strength, the magnetic ordering temperature moves to a lower value, typical for AFM materials \cite{EuMg2Bi2}. However, the ZFC and FC data exhibit a small bifurcation and the maximum in $M/H$ is rather broad. Furthermore, $M/H$ does not show a significant drop below $T_N$. These features hint at a complex incommensurate \cite{RNi2Ge2} or canted AFM structure.

Figures \ref{Fig2}(e) and \ref{Fig2}(f) display the magnetization isotherms measured at various temperatures along the two characteristic crystallographic directions. At 1.7 K, $M(H)$ initially increases linearly with field and then saturates in a high field region. The saturated value of magnetization is 7.6$\mu_B$ ($H$$\parallel c$) and 7.2$\mu_B$ ($H$$\perp c$), which is close to the theoretical value $gS\mu_B$ = 7$\mu_B$ for a Eu$^{2+}$ ion, suggesting that all the spins are aligned along the field direction. The magnetization remains nonlinear close to $T_N$. However, at high enough temperatures (\textit{T} $\ge$ 50 K) in the paramagnetic state, the magnetization shows linear variation with field. As shown in the inset of Fig. \ref{Fig2}(f) the d$M$/d$H$ vs. $H$ data taken at 1.7 K exhibit a distinct peak near 0.3 T the along $H\perp$\textit{c} axis which can be attributed to a  metamagnetic transition. The overall magnetic properties of EuAuAs are similar to those of EuCuAs \cite{EuCuAs} and EuAgAs \cite{EuAgAs}.

\subsection{Heat capacity}
\begin{figure}
	\includegraphics[width=7.9cm, keepaspectratio]{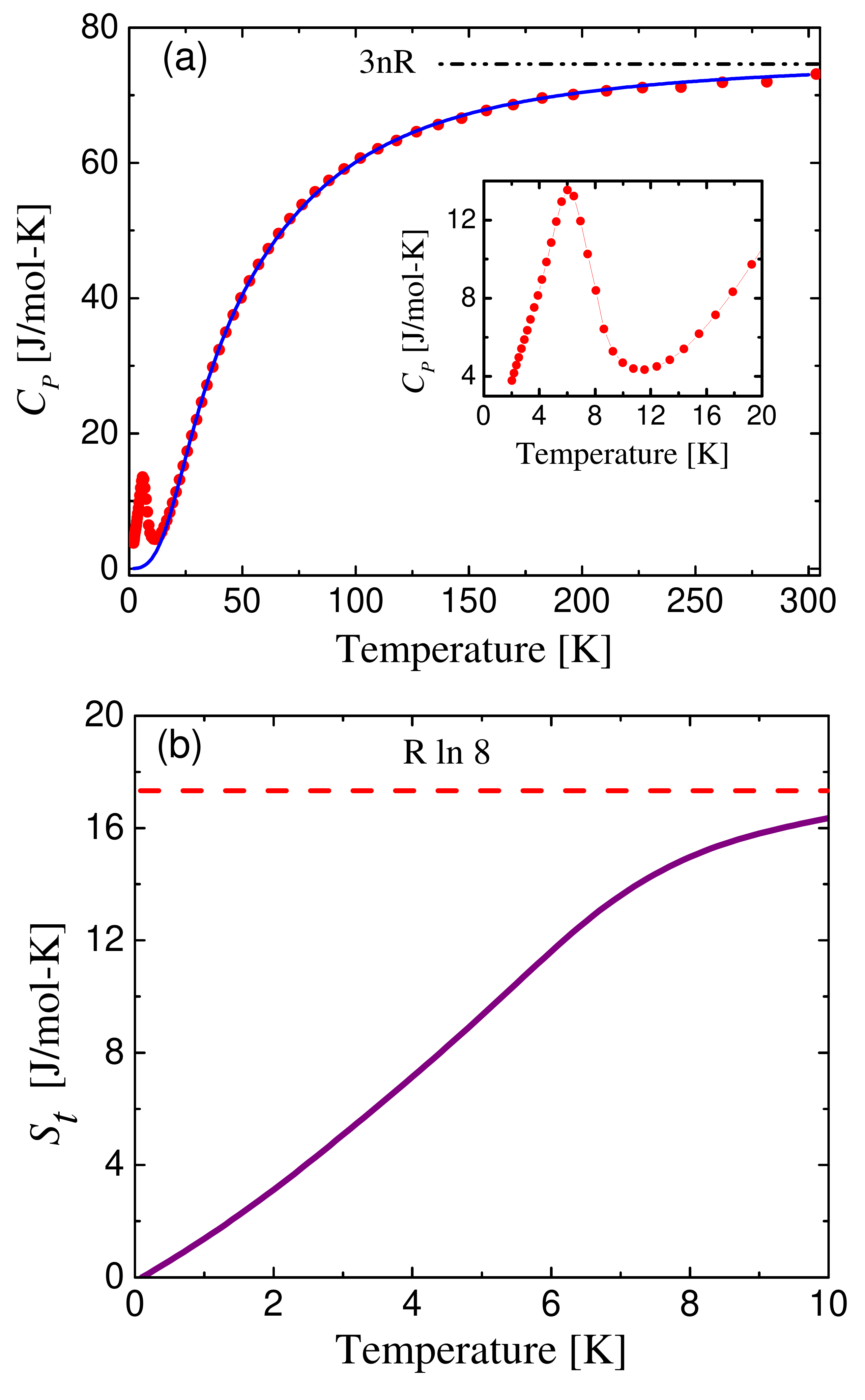}
	\caption{\label{Fig3}(a) Temperature dependence of the specific heat of EuAuAs. Solid blue line represents the fit to Eq. (\ref{Eq1}). Inset displays the zoomed view of the low-temperature data. (b) Low-temperature variation of the total entropy in EuAuAs.}
\end{figure}

The bulk nature of the magnetic ordering in EuAuAs was confirmed by the heat capacity measurement performed in zero field and constant pressure [see Fig. \ref{Fig3}(a)]. At room temperature, the value of $C_P$ (= 73.12 J/mol K) is very close to the Dulong-Petit limit $C_P=3nR$ = 74.84 J/mol K, where $n$ is the number of atoms in the formula units, and $R$ is the universal gas constant. The AFM transition manifests itself in $C_P(T)$  as a peak at $T_N$ = 6 K, in concert with the $\chi$(T) data. Interestingly, the anomaly does not show classical lambdalike shape, typical for second order magnetic transitions, but has an extended tail above $T_N$  [see the inset to Fig. \ref{Fig3}(a)]. Above 20 K, the specific heat data can be well fitted by the expression \cite{EuMg2Bi2}
\begin{equation}
	C_P(T) = \gamma T+mC_{D}+(1-m)C_{E}
	\label{Eq1}
\end{equation}
which is a sum of the electronic ($\gamma$ is the Sommerfeld coefficient), Debye and Einstein contributions (\textit{m} is the weight factor). $C_D$ and $C_E$ are defined as
\begin{equation}
	C_{D}=9nR\left( \frac{T}{\Theta_D}\right)^3\int_{0}^{\Theta_D/T}\frac{x^4e^x}{(e^x-1)^2}dx
\end{equation}

\begin{equation}
	C_{E}=3nR\left( \frac{\Theta_E}{T}\right)^2\frac{e^{\Theta_E/T}}{(e^{\Theta_E/T}-1)^2}
\end{equation}
where, $\Theta_D$ and $\Theta_E$ are the Debye and Einstein temperatures, respectively. The so-obtained values of the fitting parameters are $\gamma$ = 1.24 mJ/mol K$^2$, $\Theta_D$ = 144 K and $\Theta_E$ = 313 K and $m$ = 0.77. 
The large value of $\Theta_E$ suggests the presence of high-frequency optical modes. Since the value of the Sommerfeld coefficient obtained from the high temperature fit is unusually small for the magnetically ordered Eu based compounds, $\gamma$ was estimated also from the low temperature data. From the formula $C_P = \gamma T + \beta T^3$ applied to the data collected in the temperature range 14-22 K, the values $\gamma$ = 221 mJ/mol K$^2$ and $\Theta_D$ = 198 K were derived.  The enhanced $\gamma$ value may come from the magnon contributions, as observed in other Eu-based compounds like EuAgAs \cite{EuAgAs} and EuCr$_2$As$_2$ \cite{EuCr2As2}. Fig. \ref{Fig3}(b) displays the total entropy ($S_t$) in EuAuAs, computed as $S_t= \int \frac{C_P}{T}dT$, while the magnetic entropy expected at $T_N$ for a compound based on Eu$^{2+}$ ions is equal to $R$ln(2S+1) = $R$ln8 J/mol K. Such a value of $S_t$ is released in EuAuAs only above 10 K.

\begin{figure}
	\includegraphics[width=7.9cm, keepaspectratio]{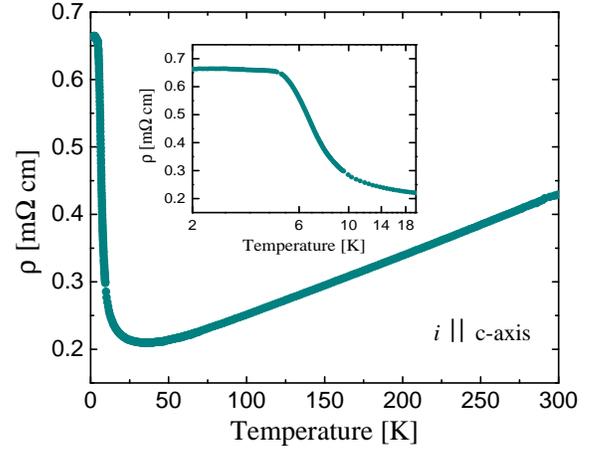}
	\caption{\label{Fig4} Temperature dependence of the electrical resistivity of EuAuAs single crystal measured with the electric current flowing along the \textit{c} axis of the hexagonal unit cell. A close view of the resistivity at low temperatures is shown in the inset (note a semilogarithmic scale).}
\end{figure}

\subsection{Magnetotransport}

The electrical resistivity of single-crystalline EuAuAs measured in zero magnetic field with electric current flowing along the \textit{c} axis of the hexagonal unit cell, $\rho(T)$, is presented in Fig. \ref{Fig4}. The room temperature resistivity value of 0.43 m$\Omega$ cm is large. Such large values were reported in other Eu-based magnetically ordered topological materials like EuMg$_2$Bi$_2$ \cite{EuMg2Bi2_JAP_2021} and EuCd$_2$As$_2$ \cite{EuCd2As2}, also in isostructural topological materials like BaAuAs and BaAgAs \cite{BaAgAs_family}. Down to about 40 K, the resistivity decreases with decreasing temperature in a metallic manner, then rapidly increases and saturates below 6 K (see also the inset to Fig. \ref{Fig4}).  The upturn in $\rho$(T) at low temperatures can be associated with an enhancement of magnetic fluctuations while the compound approaches the magnetic phase transition. The plateau in $\rho$(T) in the ordered state is an unusual effect, as the resistivity is expected to drop below the magnetic ordering temperature. This type of behavior may occur as a result of competition between bulk and surface conduction channels, observed in topological insulators such as Bi$_2$Te$_2$Se \cite{Bi2Te2Se}, YbB$_{12}$ \cite{YbB12}, and SmB$_6$ \cite{SmB6}.

\begin{figure}[t]
	\includegraphics[width=7.9cm, keepaspectratio]{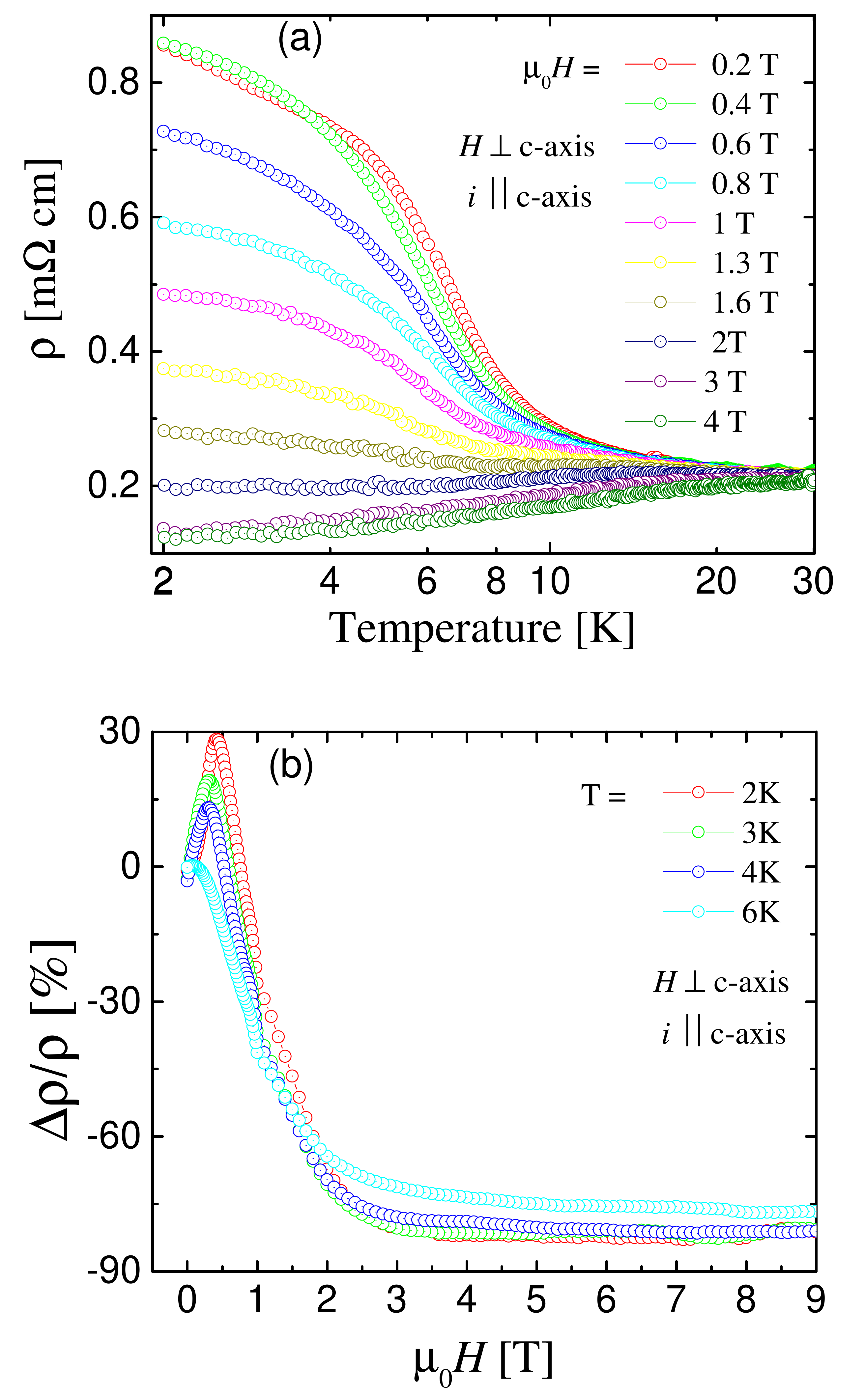}
	\caption{\label{Fig5}(a) Low-temperature electrical resistivity of single-crystalline EuAuAs measured along the \textit{c} axis in various external magnetic fields applied perpendicular to the electric current. (b) Transverse magnetoresistance isotherms of EuAuAs, measured in the AFM ordered state as specified in panel (a).}
\end{figure}

The field dependent electrical resistivity of EuAuAs, measured along the \textit{c} axis with transverse magnetic field ($H\perp$\textit{c} axis),  are presented in Fig. \ref{Fig5}(a). With the increase of magnetic field strength, the low-temperature upturn in $\rho(T)$ is getting reduced. The transverse magnetoresistance (MR), defined as  $\Delta\rho/ \rho=[\rho(H)-\rho(0)]\rho(0)$, is displayed in Fig. \ref{Fig5}(b). Initially, MR taken at 2 K increases with the increase of the applied field and reaches a maximum value of $\sim$30\% around 0.3 T. The peak in MR is associated with the metamagnetic transition as observed in the magnetization data [Fig. \ref{Fig2} (f)] . With further increase of the field, the MR value sharply decreases and becomes negative due to the suppression of magnetic fluctuations. For field values greater than 3 T, MR saturates at a large negative value of about -80\%. With increasing temperature the positive maximum in MR diminishes and shifts to lower temperatures, as expected for metamagnets, however the overall behavior of the transverse magnetoresistance isotherms remains the same. Interestingly, the saturated value of MR in the high field region hardly changes in the entire ordered state. Similar behavior of MR was reported for Eu-based antiferromagnet EuMg$_2$Bi$_2$ \cite{EuMg2Bi2_JAP_2021}  and EuBiTe$_3$ \cite{EuBiTe3}.

\subsection{Electronic structure}

\begin{figure}[t]
	\centering
	\includegraphics[width=43mm,height=40mm]{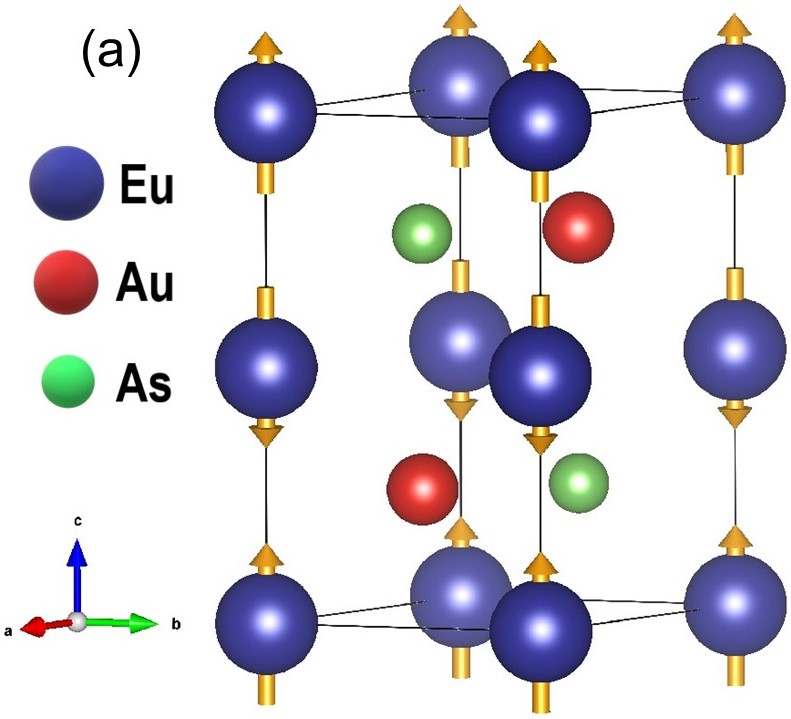}
	\hspace{0.5cm}
	\includegraphics[width=28mm,height=40mm]{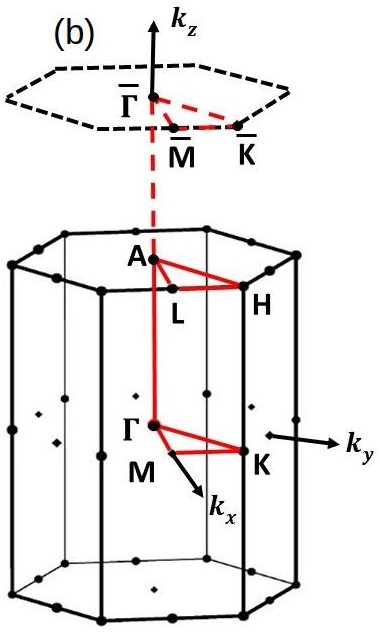}
	\caption{(a) Crystal structure of EuAuAs and (b) the irreducible Brillouin zone of the bulk along with the (001) projected surface.}
	\label{Crystal_Structure}
\end{figure}

\begin{figure}
	\centering
	\includegraphics[width=85mm,height=145mm]{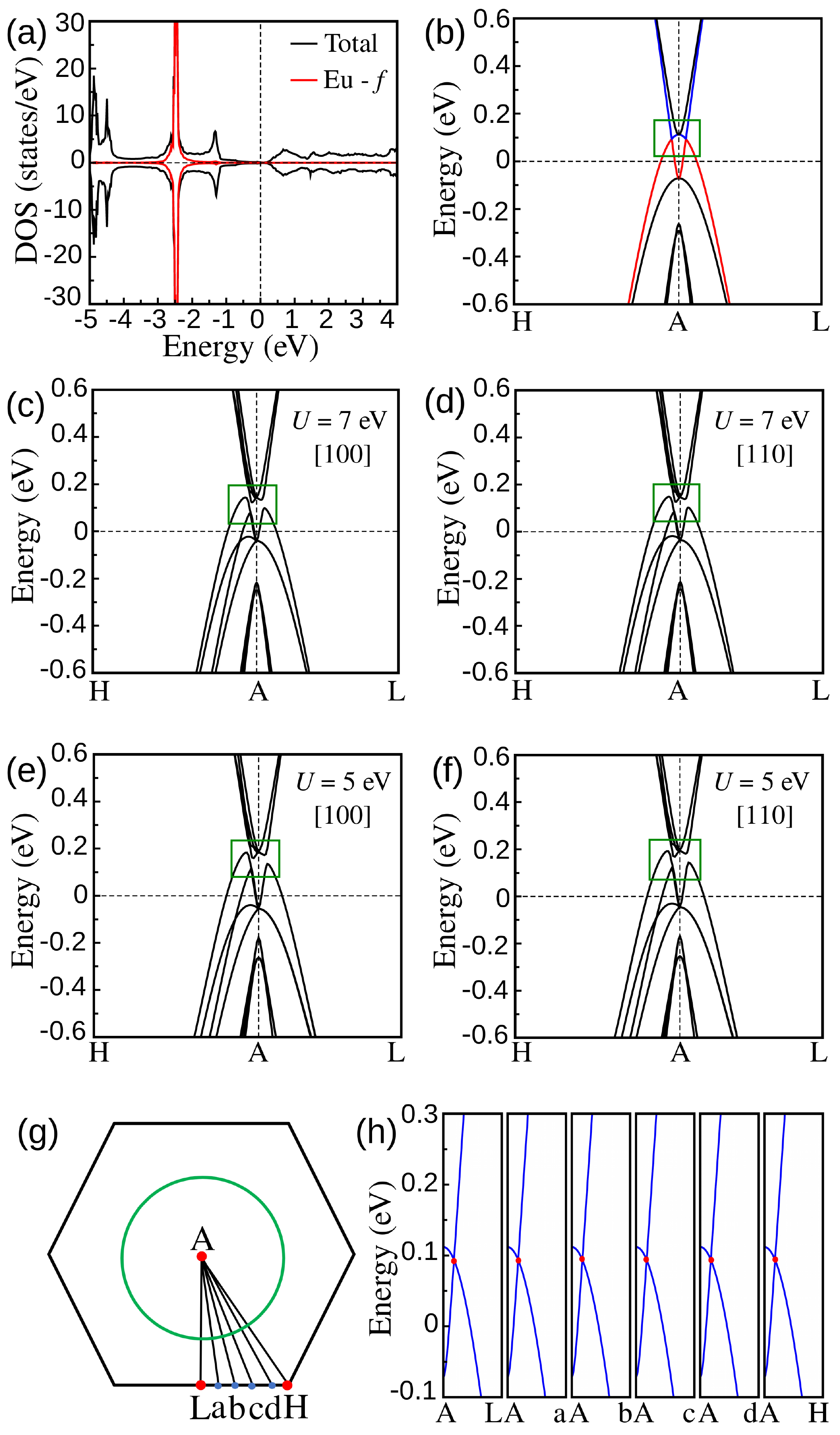}
	\caption{(a) Total and projected density of states of EuAuAs. (b) Electronic band structure along \textit{H-A-L} path without spin-orbit coupling. (c), (d) Electronic band structure along \textit{H-A-L }path at \textit{U} = 7 eV with SOC along [100] and [110] directions, respectively. (e), (f) Electronic band structure at \textit{U} = 5 eV with SOC along [100] and [110] directions, respectively. (g) The illustration of the nodal-line, where \textit{a}, \textit{b}, \textit{c} and \textit{d} are equally spaced points between \textit{L} and \textit{H}. (h) Electronic band structure at different high symmetry points [as indicated in (g)] in the Brilloiun zone in the $k_z$ = 0.5 plane.}
	\label{elec_prop}
\end{figure}

The crystal structure of EuAuAs is given in Fig. \ref{Crystal_Structure}(a). The Eu, Au, and As atoms are located at Wyckoff positions 2\textit{a}, 2\textit{d} and 2\textit{c}, respectively and form hexagonal stacking along the \textit{z} direction. The crystal structure is nonsymmorphic with preserved inversion symmetry. The ground state spin  configuration of EuAuAs was checked by calculating the total energies for possible spin configurations of AFM at U = 7 eV. We found that the system possesses FM coupling in the \textit{ab} plane and AFM coupling along \textit{c} axis i.e., it is an A-type AFM [100] system (as the ground state), and the energy difference between [100] and [110] spin configurations is very small (-5.2 $\mu$eV). We have also checked the ground state spin configuration of EuAuAs at \textit{U} = 5 eV and found the AFM [110] system to be lowest with minute total magnetic moment around 0.01$\mu_B$ which is in line with the experimental finding and previously reported EuAgAs \cite{EuAgAs}. The energy difference between [110] and [100] spin configurations for \textit{U} = 5 eV is -12 $\mu$eV which is ignorable. In addition, we have analyzed the electronic properties for U = 7 eV and 5 eV along [100] and [110] directions, and found that the electronic band structures are very similar to each other (as shown in Fig. \ref{elec_prop}). Due to the robustness of electronic properties at different U values and spin configurations, we have proceeded with AFM [100] configuration at U = 7 eV.

To demonstrate the behavior of Eu-\textit{f} states, the total density of states (DOS) and the projected density of states (PDOS) were calculated for AFM [100] at \textit{U} = 7 eV, and the results are shown in Fig. \ref{elec_prop}(a). The valence band is mainly dominated by Eu-\textit{f} states for both spin up and spin down channels and has a tiny DOS at the Fermi level. According to the electronic band structure [Fig. \ref{elec_prop}(b)], both the spin channels are conducting. The electronic band structure is very interesting with Dirac like linearized points near the Fermi level along \textit{H-A-L} path, which might further lead to a possibility of nodal-line. We prefer to focus on the red and blue bands which could be responsible for the nodal-line formation. To analyze the band-crossings along \textit{H-A-L} and to confirm the same on the plane $k_z$ = 0.5, we carefully examined the band structure along selected equally spaced paths between \textit{L} and \textit{H}, which is shown in Fig. \ref{elec_prop}(h). It can be seen that the band-crossings appear along \textit{A-a}, \textit{A-b}, \textit{A-c} and \textit{A-d} paths, which infers that an \textit{A}-centered nodal line should occur in the $k_z$ = 0.5 plane. These crossings exhibit a very little variation in energy along the selected paths, which is responsible for the flatness of the nodal line as discussed in other compounds \cite{Heja}. We conclude that these band-crossings can be classified as type-I in the $k_z$ = 0.5 plane.

\begin{figure}
	\centering
	\includegraphics[width=85mm,height=35mm]{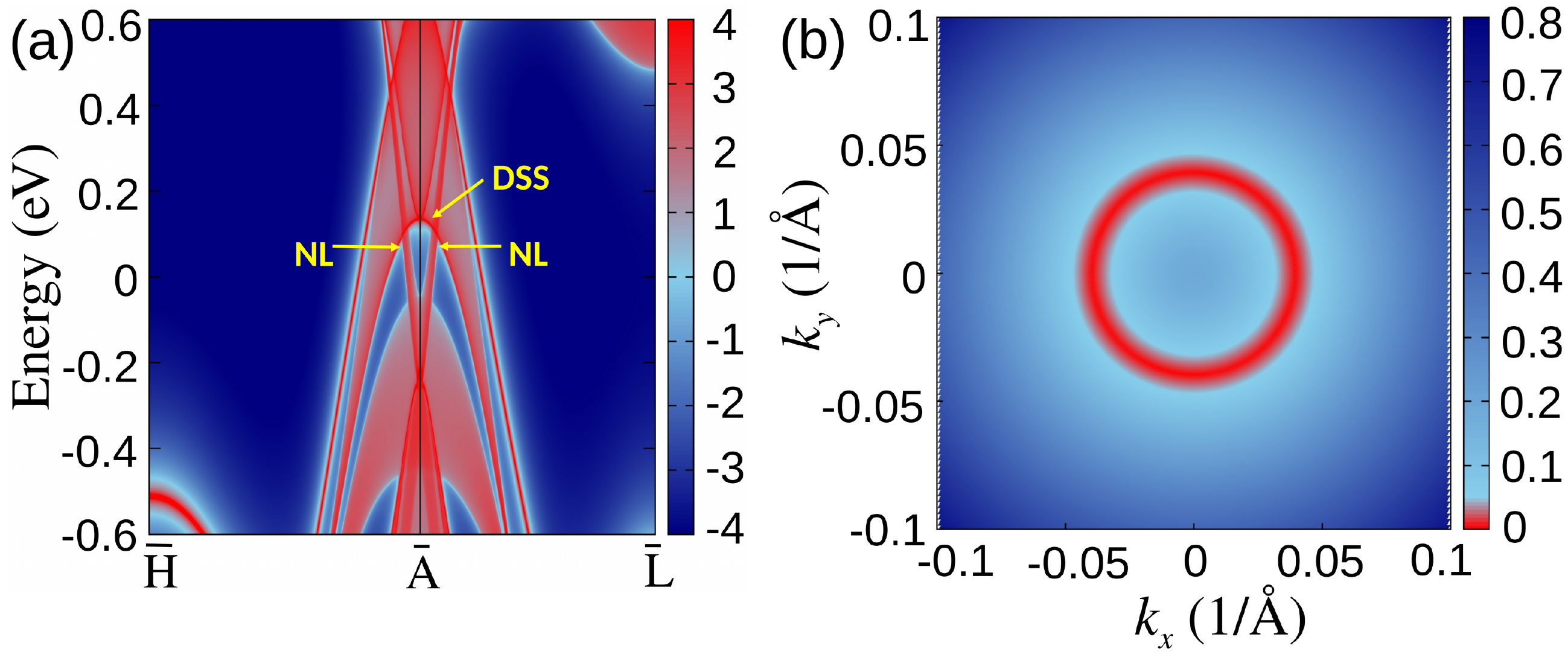}
	\caption{(a) The surface states with Eu-termination along \textit{H-A-L} path and (b) the corresponding Fermi arc in the $k_z$ = 0.5 plane.}
	\label{surface}
\end{figure}

The surface states with Eu termination are shown in Fig. \ref{surface}(a). From the plots, we can observe the nodal line slightly above the Fermi level as seen in the bulk band structure together with a drumhead surface state connecting both the crossing points along \textit{H-A} and \textit{A-L}. The gap plane calculations are performed to confirm the nodal-line which is given in Fig. \ref{surface}(b). The continuous ring seen in this figure ensures the presence of the nodal-line in the $k_z$ = 0.5 plane protected by \textit{C}$_{3z}$ rotational symmetry at high symmetry point \textit{A}.

The crossing points along with the nodal line present in the investigated system motivate us to employ spin-orbit coupling (SOC) in order to verify the nontrivial topology here. The electronic band structure with SOC is shown in Fig. \ref{elec_prop}(c). From the figure, it is vivid that two doubly degenerate bands are merging at high symmetry point \textit{A} around \textit{E} = 0.2 eV, resulting in a fourfold degenerate point. Our investigated system EuAuAs is similar to the EuAgAs \cite{EuAgAs} class of antiferromagnetic topological materials, which holds the effective time reversal symmetry as well as inversion symmetry. The presence of these symmetries along with fourfold degeneracy is sufficient to claim this point as a Dirac point, and motivate us to calculate \textit{Z}$_2$ invariants, which are (0;0 1 0). The Dirac point at high symmetry point \textit{A} and \textit{Z}$_2$ invariants are the evidences to claim EuAuAs as a weak AFM topological material.

\vspace{3mm}

\section{Conclusions}

Based on the results of thermodynamic and electrical transport measurements, performed on high-quality single crystals grown from Bi flux, we conclude that the ternary compound EuAuAs (space group $P6_3/mmc$) orders antiferromagnetically at $T_N$ = 6 K due to Eu$^{2+}$ ions. The magnetic susceptibility features hint at a complex incommensurate or canted AFM structure. In the ordered state, the magnetization first increases linearly with increasing magnetic field strength, and then saturates in high fields. The heat capacity shows a clear anomaly at $T_N$. The electrical resistivity shows an enhancement below 40 K, which is suppressed with the application of magnetic field. Below $T_N$, the transverse magnetoresistance shows a positive maximum, marking a metamagnetic transition, and in high magnetic fields it saturates at a large negative value of about 80\%. The AFM ground state in EuAuAs has been corroborated in \textit{ab initio} electronic band structure calculations. Most importantly, the theoretical computations revealed the formation of a nodal line in the bulk structure, which is placed slightly above the chemical potential and protected by rotational crystal symmetry, as well as a drumheadlike surface state, located at high symmetry point. The inclusion of spin-orbit coupling suggests that EuAuAs is a new candidate for antiferromagnetic topological material.

\section{Acknowledgment}
We thank Daloo Ram for his assistance with the crystal growth.
Financial support from the Science and Engineering Research Board (India) and 
the Department of Science and Technology (India) are acknowledged. V.K. acknowledges IIT Hyderabad for its computational facility. J.S. was supported
through CSIR scholarship. Z.H. acknowledges financial support from the Polish
National Agency for Academic Exchange under a Ulam fellowship.
	
\bibliography{Reference_EuAuAs}

\begin{thebibliography}{54}%
\makeatletter
\providecommand \@ifxundefined [1]{%
 \@ifx{#1\undefined}
}%
\providecommand \@ifnum [1]{%
 \ifnum #1\expandafter \@firstoftwo
 \else \expandafter \@secondoftwo
 \fi
}%
\providecommand \@ifx [1]{%
 \ifx #1\expandafter \@firstoftwo
 \else \expandafter \@secondoftwo
 \fi
}%
\providecommand \natexlab [1]{#1}%
\providecommand \enquote  [1]{``#1''}%
\providecommand \bibnamefont  [1]{#1}%
\providecommand \bibfnamefont [1]{#1}%
\providecommand \citenamefont [1]{#1}%
\providecommand \href@noop [0]{\@secondoftwo}%
\providecommand \href [0]{\begingroup \@sanitize@url \@href}%
\providecommand \@href[1]{\@@startlink{#1}\@@href}%
\providecommand \@@href[1]{\endgroup#1\@@endlink}%
\providecommand \@sanitize@url [0]{\catcode `\\12\catcode `\$12\catcode
  `\&12\catcode `\#12\catcode `\^12\catcode `\_12\catcode `\%12\relax}%
\providecommand \@@startlink[1]{}%
\providecommand \@@endlink[0]{}%
\providecommand \url  [0]{\begingroup\@sanitize@url \@url }%
\providecommand \@url [1]{\endgroup\@href {#1}{\urlprefix }}%
\providecommand \urlprefix  [0]{URL }%
\providecommand \Eprint [0]{\href }%
\providecommand \doibase [0]{http://dx.doi.org/}%
\providecommand \selectlanguage [0]{\@gobble}%
\providecommand \bibinfo  [0]{\@secondoftwo}%
\providecommand \bibfield  [0]{\@secondoftwo}%
\providecommand \translation [1]{[#1]}%
\providecommand \BibitemOpen [0]{}%
\providecommand \bibitemStop [0]{}%
\providecommand \bibitemNoStop [0]{.\EOS\space}%
\providecommand \EOS [0]{\spacefactor3000\relax}%
\providecommand \BibitemShut  [1]{\csname bibitem#1\endcsname}%
\let\auto@bib@innerbib\@empty
\bibitem [{\citenamefont {Jeevan}\ \emph {et~al.}(2008)\citenamefont {Jeevan},
  \citenamefont {Hossain}, \citenamefont {Kasinathan}, \citenamefont {Rosner},
  \citenamefont {Geibel},\ and\ \citenamefont {Gegenwart}}]{EuFe2As2_PRB_2008}%
  \BibitemOpen
  \bibfield  {author} {\bibinfo {author} {\bibfnamefont {H.~S.}\ \bibnamefont
  {Jeevan}}, \bibinfo {author} {\bibfnamefont {Z.}~\bibnamefont {Hossain}},
  \bibinfo {author} {\bibfnamefont {D.}~\bibnamefont {Kasinathan}}, \bibinfo
  {author} {\bibfnamefont {H.}~\bibnamefont {Rosner}}, \bibinfo {author}
  {\bibfnamefont {C.}~\bibnamefont {Geibel}}, \ and\ \bibinfo {author}
  {\bibfnamefont {P.}~\bibnamefont {Gegenwart}},\ }\href {\doibase
  10.1103/PhysRevB.78.052502} {\bibfield  {journal} {\bibinfo  {journal} {Phys.
  Rev. B}\ }\textbf {\bibinfo {volume} {78}},\ \bibinfo {pages} {052502}
  (\bibinfo {year} {2008})}\BibitemShut {NoStop}%
\bibitem [{\citenamefont {Ren}\ \emph {et~al.}(2008)\citenamefont {Ren},
  \citenamefont {Zhu}, \citenamefont {Jiang}, \citenamefont {Xu}, \citenamefont
  {Tao}, \citenamefont {Wang}, \citenamefont {Feng}, \citenamefont {Cao},\ and\
  \citenamefont {Xu}}]{EuFe2As2_PRB_2008_Possible_SC}%
  \BibitemOpen
  \bibfield  {author} {\bibinfo {author} {\bibfnamefont {Z.}~\bibnamefont
  {Ren}}, \bibinfo {author} {\bibfnamefont {Z.}~\bibnamefont {Zhu}}, \bibinfo
  {author} {\bibfnamefont {S.}~\bibnamefont {Jiang}}, \bibinfo {author}
  {\bibfnamefont {X.}~\bibnamefont {Xu}}, \bibinfo {author} {\bibfnamefont
  {Q.}~\bibnamefont {Tao}}, \bibinfo {author} {\bibfnamefont {C.}~\bibnamefont
  {Wang}}, \bibinfo {author} {\bibfnamefont {C.}~\bibnamefont {Feng}}, \bibinfo
  {author} {\bibfnamefont {G.}~\bibnamefont {Cao}}, \ and\ \bibinfo {author}
  {\bibfnamefont {Z.}~\bibnamefont {Xu}},\ }\href {\doibase
  10.1103/PhysRevB.78.052501} {\bibfield  {journal} {\bibinfo  {journal} {Phys.
  Rev. B}\ }\textbf {\bibinfo {volume} {78}},\ \bibinfo {pages} {052501}
  (\bibinfo {year} {2008})}\BibitemShut {NoStop}%
\bibitem [{\citenamefont {Jin}\ \emph {et~al.}(2019)\citenamefont {Jin},
  \citenamefont {Xiao}, \citenamefont {Nandi}, \citenamefont {Price},
  \citenamefont {Su}, \citenamefont {Schmalzl}, \citenamefont {Schmidt},
  \citenamefont {Chatterji}, \citenamefont {Thamizhavel},\ and\ \citenamefont
  {Br\"uckel}}]{EuFe2As2_PRB_2019_Pressure_Neutron}%
  \BibitemOpen
  \bibfield  {author} {\bibinfo {author} {\bibfnamefont {W.~T.}\ \bibnamefont
  {Jin}}, \bibinfo {author} {\bibfnamefont {Y.}~\bibnamefont {Xiao}}, \bibinfo
  {author} {\bibfnamefont {S.}~\bibnamefont {Nandi}}, \bibinfo {author}
  {\bibfnamefont {S.}~\bibnamefont {Price}}, \bibinfo {author} {\bibfnamefont
  {Y.}~\bibnamefont {Su}}, \bibinfo {author} {\bibfnamefont {K.}~\bibnamefont
  {Schmalzl}}, \bibinfo {author} {\bibfnamefont {W.}~\bibnamefont {Schmidt}},
  \bibinfo {author} {\bibfnamefont {T.}~\bibnamefont {Chatterji}}, \bibinfo
  {author} {\bibfnamefont {A.}~\bibnamefont {Thamizhavel}}, \ and\ \bibinfo
  {author} {\bibfnamefont {T.}~\bibnamefont {Br\"uckel}},\ }\href {\doibase
  10.1103/PhysRevB.100.014503} {\bibfield  {journal} {\bibinfo  {journal}
  {Phys. Rev. B}\ }\textbf {\bibinfo {volume} {100}},\ \bibinfo {pages}
  {014503} (\bibinfo {year} {2019})}\BibitemShut {NoStop}%
\bibitem [{\citenamefont {Anupam}\ \emph {et~al.}(2011)\citenamefont {Anupam},
  \citenamefont {Paulose}, \citenamefont {Ramakrishnan},\ and\ \citenamefont
  {Hossain}}]{EuFe2As2_JPCM_K_Doped_2011}%
  \BibitemOpen
  \bibfield  {author} {\bibinfo {author} {\bibnamefont {Anupam}}, \bibinfo
  {author} {\bibfnamefont {P.~L.}\ \bibnamefont {Paulose}}, \bibinfo {author}
  {\bibfnamefont {S.}~\bibnamefont {Ramakrishnan}}, \ and\ \bibinfo {author}
  {\bibfnamefont {Z.}~\bibnamefont {Hossain}},\ }\href {\doibase
  10.1088/0953-8984/23/45/455702} {\bibfield  {journal} {\bibinfo  {journal}
  {Journal of Physics: Condensed Matter}\ }\textbf {\bibinfo {volume} {23}},\
  \bibinfo {pages} {455702} (\bibinfo {year} {2011})}\BibitemShut {NoStop}%
\bibitem [{\citenamefont {Anupam}\ \emph {et~al.}(2012)\citenamefont {Anupam},
  \citenamefont {Anand}, \citenamefont {Paulose}, \citenamefont {Ramakrishnan},
  \citenamefont {Geibel},\ and\ \citenamefont
  {Hossain}}]{EuFe2As2_PRB_Ni_Doped_2012}%
  \BibitemOpen
  \bibfield  {author} {\bibinfo {author} {\bibnamefont {Anupam}}, \bibinfo
  {author} {\bibfnamefont {V.~K.}\ \bibnamefont {Anand}}, \bibinfo {author}
  {\bibfnamefont {P.~L.}\ \bibnamefont {Paulose}}, \bibinfo {author}
  {\bibfnamefont {S.}~\bibnamefont {Ramakrishnan}}, \bibinfo {author}
  {\bibfnamefont {C.}~\bibnamefont {Geibel}}, \ and\ \bibinfo {author}
  {\bibfnamefont {Z.}~\bibnamefont {Hossain}},\ }\href {\doibase
  10.1103/PhysRevB.85.144513} {\bibfield  {journal} {\bibinfo  {journal} {Phys.
  Rev. B}\ }\textbf {\bibinfo {volume} {85}},\ \bibinfo {pages} {144513}
  (\bibinfo {year} {2012})}\BibitemShut {NoStop}%
\bibitem [{\citenamefont {Paramanik}\ \emph
  {et~al.}(2014{\natexlab{a}})\citenamefont {Paramanik}, \citenamefont
  {Paulose}, \citenamefont {Ramakrishnan}, \citenamefont {Nigam}, \citenamefont
  {Geibel},\ and\ \citenamefont {Hossain}}]{EuFe2As2_SST_Ir_doped_2014}%
  \BibitemOpen
  \bibfield  {author} {\bibinfo {author} {\bibfnamefont {U.~B.}\ \bibnamefont
  {Paramanik}}, \bibinfo {author} {\bibfnamefont {P.~L.}\ \bibnamefont
  {Paulose}}, \bibinfo {author} {\bibfnamefont {S.}~\bibnamefont
  {Ramakrishnan}}, \bibinfo {author} {\bibfnamefont {A.~K.}\ \bibnamefont
  {Nigam}}, \bibinfo {author} {\bibfnamefont {C.}~\bibnamefont {Geibel}}, \
  and\ \bibinfo {author} {\bibfnamefont {Z.}~\bibnamefont {Hossain}},\ }\href
  {\doibase 10.1088/0953-2048/27/7/075012} {\bibfield  {journal} {\bibinfo
  {journal} {Superconductor Science and Technology}\ }\textbf {\bibinfo
  {volume} {27}},\ \bibinfo {pages} {075012} (\bibinfo {year}
  {2014}{\natexlab{a}})}\BibitemShut {NoStop}%
\bibitem [{\citenamefont {Ramarao}\ \emph {et~al.}(2020)\citenamefont
  {Ramarao}, \citenamefont {Singh}, \citenamefont {Subbarao},\ and\
  \citenamefont {Peter}}]{EuT2X2}%
  \BibitemOpen
  \bibfield  {author} {\bibinfo {author} {\bibfnamefont {S.}~\bibnamefont
  {Ramarao}}, \bibinfo {author} {\bibfnamefont {A.~K.}\ \bibnamefont {Singh}},
  \bibinfo {author} {\bibfnamefont {U.}~\bibnamefont {Subbarao}}, \ and\
  \bibinfo {author} {\bibfnamefont {S.~C.}\ \bibnamefont {Peter}},\ }\href
  {\doibase https://doi.org/10.1016/j.jssc.2019.121048} {\bibfield  {journal}
  {\bibinfo  {journal} {Journal of Solid State Chemistry}\ }\textbf {\bibinfo
  {volume} {281}},\ \bibinfo {pages} {121048} (\bibinfo {year}
  {2020})}\BibitemShut {NoStop}%
\bibitem [{\citenamefont {Pöttgen}\ and\ \citenamefont
  {Johrendt}(2000)}]{Eu_ternary_111_Pottgen}%
  \BibitemOpen
  \bibfield  {author} {\bibinfo {author} {\bibfnamefont {R.}~\bibnamefont
  {Pöttgen}}\ and\ \bibinfo {author} {\bibfnamefont {D.}~\bibnamefont
  {Johrendt}},\ }\href {\doibase 10.1021/cm991183v} {\bibfield  {journal}
  {\bibinfo  {journal} {Chemistry of Materials}\ }\textbf {\bibinfo {volume}
  {12}},\ \bibinfo {pages} {875} (\bibinfo {year} {2000})}\BibitemShut
  {NoStop}%
\bibitem [{\citenamefont {Michels}\ \emph {et~al.}(1994)\citenamefont
  {Michels}, \citenamefont {Junk}, \citenamefont {Schlabitz}, \citenamefont
  {Holland-Moritz}, \citenamefont {Abd-Elmeguid}, \citenamefont {Dunner},\ and\
  \citenamefont {Mewis}}]{Divalent_Eu}%
  \BibitemOpen
  \bibfield  {author} {\bibinfo {author} {\bibfnamefont {G.}~\bibnamefont
  {Michels}}, \bibinfo {author} {\bibfnamefont {S.}~\bibnamefont {Junk}},
  \bibinfo {author} {\bibfnamefont {W.}~\bibnamefont {Schlabitz}}, \bibinfo
  {author} {\bibfnamefont {E.}~\bibnamefont {Holland-Moritz}}, \bibinfo
  {author} {\bibfnamefont {M.~M.}\ \bibnamefont {Abd-Elmeguid}}, \bibinfo
  {author} {\bibfnamefont {J.}~\bibnamefont {Dunner}}, \ and\ \bibinfo {author}
  {\bibfnamefont {A.}~\bibnamefont {Mewis}},\ }\href {\doibase
  10.1088/0953-8984/6/9/018} {\bibfield  {journal} {\bibinfo  {journal}
  {Journal of Physics: Condensed Matter}\ }\textbf {\bibinfo {volume} {6}},\
  \bibinfo {pages} {1769} (\bibinfo {year} {1994})}\BibitemShut {NoStop}%
\bibitem [{\citenamefont {Tomuschat}\ and\ \citenamefont
  {Schuster}(1981)}]{Eu_111}%
  \BibitemOpen
  \bibfield  {author} {\bibinfo {author} {\bibfnamefont {C.}~\bibnamefont
  {Tomuschat}}\ and\ \bibinfo {author} {\bibfnamefont {H.-U.}\ \bibnamefont
  {Schuster}},\ }\href {\doibase doi:10.1515/znb-1981-0929} {\bibfield
  {journal} {\bibinfo  {journal} {Zeitschrift für Naturforschung B}\ }\textbf
  {\bibinfo {volume} {36}},\ \bibinfo {pages} {1193} (\bibinfo {year}
  {1981})}\BibitemShut {NoStop}%
\bibitem [{\citenamefont {Johrendt}\ \emph {et~al.}(1997)\citenamefont
  {Johrendt}, \citenamefont {Felser}, \citenamefont {Huhnt}, \citenamefont
  {Michels}, \citenamefont {Schäfer},\ and\ \citenamefont {Mewis}}]{EuPdAs}%
  \BibitemOpen
  \bibfield  {author} {\bibinfo {author} {\bibfnamefont {D.}~\bibnamefont
  {Johrendt}}, \bibinfo {author} {\bibfnamefont {C.}~\bibnamefont {Felser}},
  \bibinfo {author} {\bibfnamefont {C.}~\bibnamefont {Huhnt}}, \bibinfo
  {author} {\bibfnamefont {G.}~\bibnamefont {Michels}}, \bibinfo {author}
  {\bibfnamefont {W.}~\bibnamefont {Schäfer}}, \ and\ \bibinfo {author}
  {\bibfnamefont {A.}~\bibnamefont {Mewis}},\ }\href {\doibase
  https://doi.org/10.1016/S0925-8388(96)02477-2} {\bibfield  {journal}
  {\bibinfo  {journal} {Journal of Alloys and Compounds}\ }\textbf {\bibinfo
  {volume} {246}},\ \bibinfo {pages} {21} (\bibinfo {year} {1997})}\BibitemShut
  {NoStop}%
\bibitem [{\citenamefont {Ernet}\ \emph {et~al.}(1997)\citenamefont {Ernet},
  \citenamefont {Müllmann}, \citenamefont {D.~Mosel}, \citenamefont {Eckert},
  \citenamefont {Pöttgen},\ and\ \citenamefont {Kotzyba}}]{EuZnSn}%
  \BibitemOpen
  \bibfield  {author} {\bibinfo {author} {\bibfnamefont {U.}~\bibnamefont
  {Ernet}}, \bibinfo {author} {\bibfnamefont {R.}~\bibnamefont {Müllmann}},
  \bibinfo {author} {\bibfnamefont {B.}~\bibnamefont {D.~Mosel}}, \bibinfo
  {author} {\bibfnamefont {H.}~\bibnamefont {Eckert}}, \bibinfo {author}
  {\bibfnamefont {R.}~\bibnamefont {Pöttgen}}, \ and\ \bibinfo {author}
  {\bibfnamefont {G.}~\bibnamefont {Kotzyba}},\ }\href {\doibase
  10.1039/A604874I} {\bibfield  {journal} {\bibinfo  {journal} {J. Mater.
  Chem.}\ }\textbf {\bibinfo {volume} {7}},\ \bibinfo {pages} {255} (\bibinfo
  {year} {1997})}\BibitemShut {NoStop}%
\bibitem [{\citenamefont {Tong}\ \emph {et~al.}(2014)\citenamefont {Tong},
  \citenamefont {Parry}, \citenamefont {Tao}, \citenamefont {Cao},
  \citenamefont {Xu},\ and\ \citenamefont {Zeng}}]{EuCuAs}%
  \BibitemOpen
  \bibfield  {author} {\bibinfo {author} {\bibfnamefont {J.}~\bibnamefont
  {Tong}}, \bibinfo {author} {\bibfnamefont {J.}~\bibnamefont {Parry}},
  \bibinfo {author} {\bibfnamefont {Q.}~\bibnamefont {Tao}}, \bibinfo {author}
  {\bibfnamefont {G.-H.}\ \bibnamefont {Cao}}, \bibinfo {author} {\bibfnamefont
  {Z.-A.}\ \bibnamefont {Xu}}, \ and\ \bibinfo {author} {\bibfnamefont
  {H.}~\bibnamefont {Zeng}},\ }\href {\doibase
  https://doi.org/10.1016/j.jallcom.2014.02.157} {\bibfield  {journal}
  {\bibinfo  {journal} {Journal of Alloys and Compounds}\ }\textbf {\bibinfo
  {volume} {602}},\ \bibinfo {pages} {26} (\bibinfo {year} {2014})}\BibitemShut
  {NoStop}%
\bibitem [{\citenamefont {Yan}\ and\ \citenamefont
  {Felser}(2017)}]{Weyl_review}%
  \BibitemOpen
  \bibfield  {author} {\bibinfo {author} {\bibfnamefont {B.}~\bibnamefont
  {Yan}}\ and\ \bibinfo {author} {\bibfnamefont {C.}~\bibnamefont {Felser}},\
  }\href {\doibase 10.1146/annurev-conmatphys-031016-025458} {\bibfield
  {journal} {\bibinfo  {journal} {Annual Review of Condensed Matter Physics}\
  }\textbf {\bibinfo {volume} {8}},\ \bibinfo {pages} {337} (\bibinfo {year}
  {2017})}\BibitemShut {NoStop}%
\bibitem [{\citenamefont {Destraz}\ \emph {et~al.}(2020)\citenamefont
  {Destraz}, \citenamefont {Das}, \citenamefont {Tsirkin}, \citenamefont {Xu},
  \citenamefont {Neupert}, \citenamefont {Chang}, \citenamefont {Schilling},
  \citenamefont {Grushin}, \citenamefont {Kohlbrecher}, \citenamefont {Keller},
  \citenamefont {Puphal}, \citenamefont {Pomjakushina},\ and\ \citenamefont
  {White}}]{PrAlGe}%
  \BibitemOpen
  \bibfield  {author} {\bibinfo {author} {\bibfnamefont {D.}~\bibnamefont
  {Destraz}}, \bibinfo {author} {\bibfnamefont {L.}~\bibnamefont {Das}},
  \bibinfo {author} {\bibfnamefont {S.~S.}\ \bibnamefont {Tsirkin}}, \bibinfo
  {author} {\bibfnamefont {Y.}~\bibnamefont {Xu}}, \bibinfo {author}
  {\bibfnamefont {T.}~\bibnamefont {Neupert}}, \bibinfo {author} {\bibfnamefont
  {J.}~\bibnamefont {Chang}}, \bibinfo {author} {\bibfnamefont
  {A.}~\bibnamefont {Schilling}}, \bibinfo {author} {\bibfnamefont {A.~G.}\
  \bibnamefont {Grushin}}, \bibinfo {author} {\bibfnamefont {J.}~\bibnamefont
  {Kohlbrecher}}, \bibinfo {author} {\bibfnamefont {L.}~\bibnamefont {Keller}},
  \bibinfo {author} {\bibfnamefont {P.}~\bibnamefont {Puphal}}, \bibinfo
  {author} {\bibfnamefont {E.}~\bibnamefont {Pomjakushina}}, \ and\ \bibinfo
  {author} {\bibfnamefont {J.~S.}\ \bibnamefont {White}},\ }\href {\doibase
  10.1038/s41535-019-0207-7} {\bibfield  {journal} {\bibinfo  {journal} {npj
  Quantum Materials}\ }\textbf {\bibinfo {volume} {5}},\ \bibinfo {pages} {5}
  (\bibinfo {year} {2020})}\BibitemShut {NoStop}%
\bibitem [{\citenamefont {Suzuki}\ \emph {et~al.}(2016)\citenamefont {Suzuki},
  \citenamefont {Chisnell}, \citenamefont {Devarakonda}, \citenamefont {Liu},
  \citenamefont {Feng}, \citenamefont {Xiao}, \citenamefont {Lynn},\ and\
  \citenamefont {Checkelsky}}]{GdPtBi}%
  \BibitemOpen
  \bibfield  {author} {\bibinfo {author} {\bibfnamefont {T.}~\bibnamefont
  {Suzuki}}, \bibinfo {author} {\bibfnamefont {R.}~\bibnamefont {Chisnell}},
  \bibinfo {author} {\bibfnamefont {A.}~\bibnamefont {Devarakonda}}, \bibinfo
  {author} {\bibfnamefont {Y.-T.}\ \bibnamefont {Liu}}, \bibinfo {author}
  {\bibfnamefont {W.}~\bibnamefont {Feng}}, \bibinfo {author} {\bibfnamefont
  {D.}~\bibnamefont {Xiao}}, \bibinfo {author} {\bibfnamefont {J.~.~W.}\
  \bibnamefont {Lynn}}, \ and\ \bibinfo {author} {\bibfnamefont {J.~.~G.}\
  \bibnamefont {Checkelsky}},\ }\href {\doibase 10.1038/nphys3831} {\bibfield
  {journal} {\bibinfo  {journal} {Nature Physics}\ }\textbf {\bibinfo {volume}
  {12}},\ \bibinfo {pages} {1119} (\bibinfo {year} {2016})}\BibitemShut
  {NoStop}%
\bibitem [{\citenamefont {Liu}\ \emph {et~al.}(2018)\citenamefont {Liu},
  \citenamefont {Sun}, \citenamefont {Kumar}, \citenamefont {Muechler},
  \citenamefont {Sun}, \citenamefont {Jiao}, \citenamefont {Yang},
  \citenamefont {Liu}, \citenamefont {Liang}, \citenamefont {Xu}, \citenamefont
  {Kroder}, \citenamefont {S{\"u}{\ss}}, \citenamefont {Borrmann},
  \citenamefont {Shekhar}, \citenamefont {Wang}, \citenamefont {Xi},
  \citenamefont {Wang}, \citenamefont {Schnelle}, \citenamefont {Wirth},
  \citenamefont {Chen}, \citenamefont {Goennenwein},\ and\ \citenamefont
  {Felser}}]{Co3SnS2}%
  \BibitemOpen
  \bibfield  {author} {\bibinfo {author} {\bibfnamefont {E.}~\bibnamefont
  {Liu}}, \bibinfo {author} {\bibfnamefont {Y.}~\bibnamefont {Sun}}, \bibinfo
  {author} {\bibfnamefont {N.}~\bibnamefont {Kumar}}, \bibinfo {author}
  {\bibfnamefont {L.}~\bibnamefont {Muechler}}, \bibinfo {author}
  {\bibfnamefont {A.}~\bibnamefont {Sun}}, \bibinfo {author} {\bibfnamefont
  {L.}~\bibnamefont {Jiao}}, \bibinfo {author} {\bibfnamefont {S.-Y.}\
  \bibnamefont {Yang}}, \bibinfo {author} {\bibfnamefont {D.}~\bibnamefont
  {Liu}}, \bibinfo {author} {\bibfnamefont {A.}~\bibnamefont {Liang}}, \bibinfo
  {author} {\bibfnamefont {Q.}~\bibnamefont {Xu}}, \bibinfo {author}
  {\bibfnamefont {J.}~\bibnamefont {Kroder}}, \bibinfo {author} {\bibfnamefont
  {V.}~\bibnamefont {S{\"u}{\ss}}}, \bibinfo {author} {\bibfnamefont
  {H.}~\bibnamefont {Borrmann}}, \bibinfo {author} {\bibfnamefont
  {C.}~\bibnamefont {Shekhar}}, \bibinfo {author} {\bibfnamefont
  {Z.}~\bibnamefont {Wang}}, \bibinfo {author} {\bibfnamefont {C.}~\bibnamefont
  {Xi}}, \bibinfo {author} {\bibfnamefont {W.}~\bibnamefont {Wang}}, \bibinfo
  {author} {\bibfnamefont {W.}~\bibnamefont {Schnelle}}, \bibinfo {author}
  {\bibfnamefont {S.}~\bibnamefont {Wirth}}, \bibinfo {author} {\bibfnamefont
  {Y.}~\bibnamefont {Chen}}, \bibinfo {author} {\bibfnamefont {S.~T.~B.}\
  \bibnamefont {Goennenwein}}, \ and\ \bibinfo {author} {\bibfnamefont
  {C.}~\bibnamefont {Felser}},\ }\href {\doibase 10.1038/s41567-018-0234-5}
  {\bibfield  {journal} {\bibinfo  {journal} {Nature Physics}\ }\textbf
  {\bibinfo {volume} {14}},\ \bibinfo {pages} {1125} (\bibinfo {year}
  {2018})}\BibitemShut {NoStop}%
\bibitem [{\citenamefont {Sakai}\ \emph {et~al.}(2018)\citenamefont {Sakai},
  \citenamefont {Mizuta}, \citenamefont {Nugroho}, \citenamefont {Sihombing},
  \citenamefont {Koretsune}, \citenamefont {Suzuki}, \citenamefont {Takemori},
  \citenamefont {Ishii}, \citenamefont {Nishio-Hamane}, \citenamefont {Arita},
  \citenamefont {Goswami},\ and\ \citenamefont {Nakatsuji}}]{Co2MnGa}%
  \BibitemOpen
  \bibfield  {author} {\bibinfo {author} {\bibfnamefont {A.}~\bibnamefont
  {Sakai}}, \bibinfo {author} {\bibfnamefont {Y.~P.}\ \bibnamefont {Mizuta}},
  \bibinfo {author} {\bibfnamefont {A.~A.}\ \bibnamefont {Nugroho}}, \bibinfo
  {author} {\bibfnamefont {R.}~\bibnamefont {Sihombing}}, \bibinfo {author}
  {\bibfnamefont {T.}~\bibnamefont {Koretsune}}, \bibinfo {author}
  {\bibfnamefont {M.-T.}\ \bibnamefont {Suzuki}}, \bibinfo {author}
  {\bibfnamefont {N.}~\bibnamefont {Takemori}}, \bibinfo {author}
  {\bibfnamefont {R.}~\bibnamefont {Ishii}}, \bibinfo {author} {\bibfnamefont
  {D.}~\bibnamefont {Nishio-Hamane}}, \bibinfo {author} {\bibfnamefont
  {R.}~\bibnamefont {Arita}}, \bibinfo {author} {\bibfnamefont
  {P.}~\bibnamefont {Goswami}}, \ and\ \bibinfo {author} {\bibfnamefont
  {S.}~\bibnamefont {Nakatsuji}},\ }\href {\doibase 10.1038/s41567-018-0225-6}
  {\bibfield  {journal} {\bibinfo  {journal} {Nature Physics}\ }\textbf
  {\bibinfo {volume} {14}},\ \bibinfo {pages} {1119} (\bibinfo {year}
  {2018})}\BibitemShut {NoStop}%
\bibitem [{\citenamefont {Guin}\ \emph {et~al.}(2019)\citenamefont {Guin},
  \citenamefont {Manna}, \citenamefont {Noky}, \citenamefont {Watzman},
  \citenamefont {Fu}, \citenamefont {Kumar}, \citenamefont {Schnelle},
  \citenamefont {Shekhar}, \citenamefont {Sun}, \citenamefont {Gooth},\ and\
  \citenamefont {Felser}}]{Co2MnGa_2019}%
  \BibitemOpen
  \bibfield  {author} {\bibinfo {author} {\bibfnamefont {S.~N.}\ \bibnamefont
  {Guin}}, \bibinfo {author} {\bibfnamefont {K.}~\bibnamefont {Manna}},
  \bibinfo {author} {\bibfnamefont {J.}~\bibnamefont {Noky}}, \bibinfo {author}
  {\bibfnamefont {S.~J.}\ \bibnamefont {Watzman}}, \bibinfo {author}
  {\bibfnamefont {C.}~\bibnamefont {Fu}}, \bibinfo {author} {\bibfnamefont
  {N.}~\bibnamefont {Kumar}}, \bibinfo {author} {\bibfnamefont
  {W.}~\bibnamefont {Schnelle}}, \bibinfo {author} {\bibfnamefont
  {C.}~\bibnamefont {Shekhar}}, \bibinfo {author} {\bibfnamefont
  {Y.}~\bibnamefont {Sun}}, \bibinfo {author} {\bibfnamefont {J.}~\bibnamefont
  {Gooth}}, \ and\ \bibinfo {author} {\bibfnamefont {C.}~\bibnamefont
  {Felser}},\ }\href {\doibase 10.1038/s41427-019-0116-z} {\bibfield  {journal}
  {\bibinfo  {journal} {NPG Asia Materials}\ }\textbf {\bibinfo {volume}
  {11}},\ \bibinfo {pages} {16} (\bibinfo {year} {2019})}\BibitemShut {NoStop}%
\bibitem [{\citenamefont {Kim}\ \emph {et~al.}(2018)\citenamefont {Kim},
  \citenamefont {Seo}, \citenamefont {Lee}, \citenamefont {Ko}, \citenamefont
  {Kim}, \citenamefont {Jang}, \citenamefont {Ok}, \citenamefont {Lee},
  \citenamefont {Jo}, \citenamefont {Kang}, \citenamefont {Shim}, \citenamefont
  {Kim}, \citenamefont {Yeom}, \citenamefont {Il~Min}, \citenamefont {Yang},\
  and\ \citenamefont {Kim}}]{Fe3GeTe2}%
  \BibitemOpen
  \bibfield  {author} {\bibinfo {author} {\bibfnamefont {K.}~\bibnamefont
  {Kim}}, \bibinfo {author} {\bibfnamefont {J.}~\bibnamefont {Seo}}, \bibinfo
  {author} {\bibfnamefont {E.}~\bibnamefont {Lee}}, \bibinfo {author}
  {\bibfnamefont {K.-T.}\ \bibnamefont {Ko}}, \bibinfo {author} {\bibfnamefont
  {B.~S.}\ \bibnamefont {Kim}}, \bibinfo {author} {\bibfnamefont {B.~G.}\
  \bibnamefont {Jang}}, \bibinfo {author} {\bibfnamefont {J.~M.}\ \bibnamefont
  {Ok}}, \bibinfo {author} {\bibfnamefont {J.}~\bibnamefont {Lee}}, \bibinfo
  {author} {\bibfnamefont {Y.~J.}\ \bibnamefont {Jo}}, \bibinfo {author}
  {\bibfnamefont {W.}~\bibnamefont {Kang}}, \bibinfo {author} {\bibfnamefont
  {J.~H.}\ \bibnamefont {Shim}}, \bibinfo {author} {\bibfnamefont
  {C.}~\bibnamefont {Kim}}, \bibinfo {author} {\bibfnamefont {H.~W.}\
  \bibnamefont {Yeom}}, \bibinfo {author} {\bibfnamefont {B.}~\bibnamefont
  {Il~Min}}, \bibinfo {author} {\bibfnamefont {B.-J.}\ \bibnamefont {Yang}}, \
  and\ \bibinfo {author} {\bibfnamefont {J.~S.}\ \bibnamefont {Kim}},\ }\href
  {\doibase 10.1038/s41563-018-0132-3} {\bibfield  {journal} {\bibinfo
  {journal} {Nature Materials}\ }\textbf {\bibinfo {volume} {17}},\ \bibinfo
  {pages} {794} (\bibinfo {year} {2018})}\BibitemShut {NoStop}%
\bibitem [{\citenamefont {Nakatsuji}\ \emph {et~al.}(2015)\citenamefont
  {Nakatsuji}, \citenamefont {Kiyohara},\ and\ \citenamefont
  {Higo}}]{Mn3Sn_Nature_2015}%
  \BibitemOpen
  \bibfield  {author} {\bibinfo {author} {\bibfnamefont {S.}~\bibnamefont
  {Nakatsuji}}, \bibinfo {author} {\bibfnamefont {N.}~\bibnamefont {Kiyohara}},
  \ and\ \bibinfo {author} {\bibfnamefont {T.}~\bibnamefont {Higo}},\ }\href
  {\doibase 10.1038/nature15723} {\bibfield  {journal} {\bibinfo  {journal}
  {Nature}\ }\textbf {\bibinfo {volume} {527}},\ \bibinfo {pages} {212}
  (\bibinfo {year} {2015})}\BibitemShut {NoStop}%
\bibitem [{\citenamefont {Chang}\ \emph {et~al.}(2018)\citenamefont {Chang},
  \citenamefont {Singh}, \citenamefont {Xu}, \citenamefont {Bian},
  \citenamefont {Huang}, \citenamefont {Hsu}, \citenamefont {Belopolski},
  \citenamefont {Alidoust}, \citenamefont {Sanchez}, \citenamefont {Zheng},
  \citenamefont {Lu}, \citenamefont {Zhang}, \citenamefont {Bian},
  \citenamefont {Chang}, \citenamefont {Jeng}, \citenamefont {Bansil},
  \citenamefont {Hsu}, \citenamefont {Jia}, \citenamefont {Neupert},
  \citenamefont {Lin},\ and\ \citenamefont {Hasan}}]{RAlGe}%
  \BibitemOpen
  \bibfield  {author} {\bibinfo {author} {\bibfnamefont {G.}~\bibnamefont
  {Chang}}, \bibinfo {author} {\bibfnamefont {B.}~\bibnamefont {Singh}},
  \bibinfo {author} {\bibfnamefont {S.-Y.}\ \bibnamefont {Xu}}, \bibinfo
  {author} {\bibfnamefont {G.}~\bibnamefont {Bian}}, \bibinfo {author}
  {\bibfnamefont {S.-M.}\ \bibnamefont {Huang}}, \bibinfo {author}
  {\bibfnamefont {C.-H.}\ \bibnamefont {Hsu}}, \bibinfo {author} {\bibfnamefont
  {I.}~\bibnamefont {Belopolski}}, \bibinfo {author} {\bibfnamefont
  {N.}~\bibnamefont {Alidoust}}, \bibinfo {author} {\bibfnamefont {D.~S.}\
  \bibnamefont {Sanchez}}, \bibinfo {author} {\bibfnamefont {H.}~\bibnamefont
  {Zheng}}, \bibinfo {author} {\bibfnamefont {H.}~\bibnamefont {Lu}}, \bibinfo
  {author} {\bibfnamefont {X.}~\bibnamefont {Zhang}}, \bibinfo {author}
  {\bibfnamefont {Y.}~\bibnamefont {Bian}}, \bibinfo {author} {\bibfnamefont
  {T.-R.}\ \bibnamefont {Chang}}, \bibinfo {author} {\bibfnamefont {H.-T.}\
  \bibnamefont {Jeng}}, \bibinfo {author} {\bibfnamefont {A.}~\bibnamefont
  {Bansil}}, \bibinfo {author} {\bibfnamefont {H.}~\bibnamefont {Hsu}},
  \bibinfo {author} {\bibfnamefont {S.}~\bibnamefont {Jia}}, \bibinfo {author}
  {\bibfnamefont {T.}~\bibnamefont {Neupert}}, \bibinfo {author} {\bibfnamefont
  {H.}~\bibnamefont {Lin}}, \ and\ \bibinfo {author} {\bibfnamefont {M.~Z.}\
  \bibnamefont {Hasan}},\ }\href {\doibase 10.1103/PhysRevB.97.041104}
  {\bibfield  {journal} {\bibinfo  {journal} {Phys. Rev. B}\ }\textbf {\bibinfo
  {volume} {97}},\ \bibinfo {pages} {041104} (\bibinfo {year}
  {2018})}\BibitemShut {NoStop}%
\bibitem [{\citenamefont {Puphal}\ \emph {et~al.}(2020)\citenamefont {Puphal},
  \citenamefont {Pomjakushin}, \citenamefont {Kanazawa}, \citenamefont
  {Ukleev}, \citenamefont {Gawryluk}, \citenamefont {Ma}, \citenamefont
  {Naamneh}, \citenamefont {Plumb}, \citenamefont {Keller}, \citenamefont
  {Cubitt}, \citenamefont {Pomjakushina},\ and\ \citenamefont
  {White}}]{CeAlGe}%
  \BibitemOpen
  \bibfield  {author} {\bibinfo {author} {\bibfnamefont {P.}~\bibnamefont
  {Puphal}}, \bibinfo {author} {\bibfnamefont {V.}~\bibnamefont {Pomjakushin}},
  \bibinfo {author} {\bibfnamefont {N.}~\bibnamefont {Kanazawa}}, \bibinfo
  {author} {\bibfnamefont {V.}~\bibnamefont {Ukleev}}, \bibinfo {author}
  {\bibfnamefont {D.~J.}\ \bibnamefont {Gawryluk}}, \bibinfo {author}
  {\bibfnamefont {J.}~\bibnamefont {Ma}}, \bibinfo {author} {\bibfnamefont
  {M.}~\bibnamefont {Naamneh}}, \bibinfo {author} {\bibfnamefont {N.~C.}\
  \bibnamefont {Plumb}}, \bibinfo {author} {\bibfnamefont {L.}~\bibnamefont
  {Keller}}, \bibinfo {author} {\bibfnamefont {R.}~\bibnamefont {Cubitt}},
  \bibinfo {author} {\bibfnamefont {E.}~\bibnamefont {Pomjakushina}}, \ and\
  \bibinfo {author} {\bibfnamefont {J.~S.}\ \bibnamefont {White}},\ }\href
  {\doibase 10.1103/PhysRevLett.124.017202} {\bibfield  {journal} {\bibinfo
  {journal} {Phys. Rev. Lett.}\ }\textbf {\bibinfo {volume} {124}},\ \bibinfo
  {pages} {017202} (\bibinfo {year} {2020})}\BibitemShut {NoStop}%
\bibitem [{\citenamefont {Rout}\ \emph {et~al.}(2019)\citenamefont {Rout},
  \citenamefont {Madduri}, \citenamefont {Manna},\ and\ \citenamefont
  {Nayak}}]{Mn3Sn}%
  \BibitemOpen
  \bibfield  {author} {\bibinfo {author} {\bibfnamefont {P.~K.}\ \bibnamefont
  {Rout}}, \bibinfo {author} {\bibfnamefont {P.~V.~P.}\ \bibnamefont
  {Madduri}}, \bibinfo {author} {\bibfnamefont {S.~K.}\ \bibnamefont {Manna}},
  \ and\ \bibinfo {author} {\bibfnamefont {A.~K.}\ \bibnamefont {Nayak}},\
  }\href {\doibase 10.1103/PhysRevB.99.094430} {\bibfield  {journal} {\bibinfo
  {journal} {Phys. Rev. B}\ }\textbf {\bibinfo {volume} {99}},\ \bibinfo
  {pages} {094430} (\bibinfo {year} {2019})}\BibitemShut {NoStop}%
\bibitem [{\citenamefont {Mardanya}\ \emph {et~al.}(2019)\citenamefont
  {Mardanya}, \citenamefont {Singh}, \citenamefont {Huang}, \citenamefont
  {Chang}, \citenamefont {Su}, \citenamefont {Lin}, \citenamefont {Agarwal},\
  and\ \citenamefont {Bansil}}]{BaAgAs}%
  \BibitemOpen
  \bibfield  {author} {\bibinfo {author} {\bibfnamefont {S.}~\bibnamefont
  {Mardanya}}, \bibinfo {author} {\bibfnamefont {B.}~\bibnamefont {Singh}},
  \bibinfo {author} {\bibfnamefont {S.-M.}\ \bibnamefont {Huang}}, \bibinfo
  {author} {\bibfnamefont {T.-R.}\ \bibnamefont {Chang}}, \bibinfo {author}
  {\bibfnamefont {C.}~\bibnamefont {Su}}, \bibinfo {author} {\bibfnamefont
  {H.}~\bibnamefont {Lin}}, \bibinfo {author} {\bibfnamefont {A.}~\bibnamefont
  {Agarwal}}, \ and\ \bibinfo {author} {\bibfnamefont {A.}~\bibnamefont
  {Bansil}},\ }\href {\doibase 10.1103/PhysRevMaterials.3.071201} {\bibfield
  {journal} {\bibinfo  {journal} {Phys. Rev. Materials}\ }\textbf {\bibinfo
  {volume} {3}},\ \bibinfo {pages} {071201} (\bibinfo {year}
  {2019})}\BibitemShut {NoStop}%
\bibitem [{\citenamefont {Mondal}\ \emph {et~al.}(2019)\citenamefont {Mondal},
  \citenamefont {Barman}, \citenamefont {Alam},\ and\ \citenamefont
  {Pathak}}]{SrAgAs}%
  \BibitemOpen
  \bibfield  {author} {\bibinfo {author} {\bibfnamefont {C.}~\bibnamefont
  {Mondal}}, \bibinfo {author} {\bibfnamefont {C.~K.}\ \bibnamefont {Barman}},
  \bibinfo {author} {\bibfnamefont {A.}~\bibnamefont {Alam}}, \ and\ \bibinfo
  {author} {\bibfnamefont {B.}~\bibnamefont {Pathak}},\ }\href {\doibase
  10.1103/PhysRevB.99.205112} {\bibfield  {journal} {\bibinfo  {journal} {Phys.
  Rev. B}\ }\textbf {\bibinfo {volume} {99}},\ \bibinfo {pages} {205112}
  (\bibinfo {year} {2019})}\BibitemShut {NoStop}%
\bibitem [{\citenamefont {Laha}\ \emph {et~al.}(2021)\citenamefont {Laha},
  \citenamefont {Singha}, \citenamefont {Mardanya}, \citenamefont {Singh},
  \citenamefont {Agarwal}, \citenamefont {Mandal},\ and\ \citenamefont
  {Hossain}}]{EuAgAs}%
  \BibitemOpen
  \bibfield  {author} {\bibinfo {author} {\bibfnamefont {A.}~\bibnamefont
  {Laha}}, \bibinfo {author} {\bibfnamefont {R.}~\bibnamefont {Singha}},
  \bibinfo {author} {\bibfnamefont {S.}~\bibnamefont {Mardanya}}, \bibinfo
  {author} {\bibfnamefont {B.}~\bibnamefont {Singh}}, \bibinfo {author}
  {\bibfnamefont {A.}~\bibnamefont {Agarwal}}, \bibinfo {author} {\bibfnamefont
  {P.}~\bibnamefont {Mandal}}, \ and\ \bibinfo {author} {\bibfnamefont
  {Z.}~\bibnamefont {Hossain}},\ }\href {\doibase 10.1103/PhysRevB.103.L241112}
  {\bibfield  {journal} {\bibinfo  {journal} {Phys. Rev. B}\ }\textbf {\bibinfo
  {volume} {103}},\ \bibinfo {pages} {L241112} (\bibinfo {year}
  {2021})}\BibitemShut {NoStop}%
\bibitem [{\citenamefont {Nakayama}\ \emph {et~al.}(2020)\citenamefont
  {Nakayama}, \citenamefont {Wang}, \citenamefont {Takane}, \citenamefont
  {Souma}, \citenamefont {Kubota}, \citenamefont {Nakata}, \citenamefont
  {Cacho}, \citenamefont {Kim}, \citenamefont {Ekahana}, \citenamefont {Shi},
  \citenamefont {Kitamura}, \citenamefont {Horiba}, \citenamefont
  {Kumigashira}, \citenamefont {Takahashi}, \citenamefont {Ando},\ and\
  \citenamefont {Sato}}]{CaAuAs_ARPES}%
  \BibitemOpen
  \bibfield  {author} {\bibinfo {author} {\bibfnamefont {K.}~\bibnamefont
  {Nakayama}}, \bibinfo {author} {\bibfnamefont {Z.}~\bibnamefont {Wang}},
  \bibinfo {author} {\bibfnamefont {D.}~\bibnamefont {Takane}}, \bibinfo
  {author} {\bibfnamefont {S.}~\bibnamefont {Souma}}, \bibinfo {author}
  {\bibfnamefont {Y.}~\bibnamefont {Kubota}}, \bibinfo {author} {\bibfnamefont
  {Y.}~\bibnamefont {Nakata}}, \bibinfo {author} {\bibfnamefont
  {C.}~\bibnamefont {Cacho}}, \bibinfo {author} {\bibfnamefont
  {T.}~\bibnamefont {Kim}}, \bibinfo {author} {\bibfnamefont {S.~A.}\
  \bibnamefont {Ekahana}}, \bibinfo {author} {\bibfnamefont {M.}~\bibnamefont
  {Shi}}, \bibinfo {author} {\bibfnamefont {M.}~\bibnamefont {Kitamura}},
  \bibinfo {author} {\bibfnamefont {K.}~\bibnamefont {Horiba}}, \bibinfo
  {author} {\bibfnamefont {H.}~\bibnamefont {Kumigashira}}, \bibinfo {author}
  {\bibfnamefont {T.}~\bibnamefont {Takahashi}}, \bibinfo {author}
  {\bibfnamefont {Y.}~\bibnamefont {Ando}}, \ and\ \bibinfo {author}
  {\bibfnamefont {T.}~\bibnamefont {Sato}},\ }\href {\doibase
  10.1103/PhysRevB.102.041104} {\bibfield  {journal} {\bibinfo  {journal}
  {Phys. Rev. B}\ }\textbf {\bibinfo {volume} {102}},\ \bibinfo {pages}
  {041104} (\bibinfo {year} {2020})}\BibitemShut {NoStop}%
\bibitem [{\citenamefont {Emmanouilidou}\ \emph {et~al.}(2017)\citenamefont
  {Emmanouilidou}, \citenamefont {Shen}, \citenamefont {Deng}, \citenamefont
  {Chang}, \citenamefont {Shi}, \citenamefont {Kotliar}, \citenamefont {Xu},\
  and\ \citenamefont {Ni}}]{CaTX}%
  \BibitemOpen
  \bibfield  {author} {\bibinfo {author} {\bibfnamefont {E.}~\bibnamefont
  {Emmanouilidou}}, \bibinfo {author} {\bibfnamefont {B.}~\bibnamefont {Shen}},
  \bibinfo {author} {\bibfnamefont {X.}~\bibnamefont {Deng}}, \bibinfo {author}
  {\bibfnamefont {T.-R.}\ \bibnamefont {Chang}}, \bibinfo {author}
  {\bibfnamefont {A.}~\bibnamefont {Shi}}, \bibinfo {author} {\bibfnamefont
  {G.}~\bibnamefont {Kotliar}}, \bibinfo {author} {\bibfnamefont {S.-Y.}\
  \bibnamefont {Xu}}, \ and\ \bibinfo {author} {\bibfnamefont {N.}~\bibnamefont
  {Ni}},\ }\href {\doibase 10.1103/PhysRevB.95.245113} {\bibfield  {journal}
  {\bibinfo  {journal} {Phys. Rev. B}\ }\textbf {\bibinfo {volume} {95}},\
  \bibinfo {pages} {245113} (\bibinfo {year} {2017})}\BibitemShut {NoStop}%
\bibitem [{\citenamefont {Laha}\ \emph
  {et~al.}(2020{\natexlab{a}})\citenamefont {Laha}, \citenamefont {Mardanya},
  \citenamefont {Singh}, \citenamefont {Lin}, \citenamefont {Bansil},
  \citenamefont {Agarwal},\ and\ \citenamefont {Hossain}}]{CaCdSn}%
  \BibitemOpen
  \bibfield  {author} {\bibinfo {author} {\bibfnamefont {A.}~\bibnamefont
  {Laha}}, \bibinfo {author} {\bibfnamefont {S.}~\bibnamefont {Mardanya}},
  \bibinfo {author} {\bibfnamefont {B.}~\bibnamefont {Singh}}, \bibinfo
  {author} {\bibfnamefont {H.}~\bibnamefont {Lin}}, \bibinfo {author}
  {\bibfnamefont {A.}~\bibnamefont {Bansil}}, \bibinfo {author} {\bibfnamefont
  {A.}~\bibnamefont {Agarwal}}, \ and\ \bibinfo {author} {\bibfnamefont
  {Z.}~\bibnamefont {Hossain}},\ }\href {\doibase 10.1103/PhysRevB.102.035164}
  {\bibfield  {journal} {\bibinfo  {journal} {Phys. Rev. B}\ }\textbf {\bibinfo
  {volume} {102}},\ \bibinfo {pages} {035164} (\bibinfo {year}
  {2020}{\natexlab{a}})}\BibitemShut {NoStop}%
\bibitem [{\citenamefont {Laha}\ \emph {et~al.}(2019)\citenamefont {Laha},
  \citenamefont {Malick}, \citenamefont {Singha}, \citenamefont {Mandal},
  \citenamefont {Rambabu}, \citenamefont {Kanchana},\ and\ \citenamefont
  {Hossain}}]{YbCdGe}%
  \BibitemOpen
  \bibfield  {author} {\bibinfo {author} {\bibfnamefont {A.}~\bibnamefont
  {Laha}}, \bibinfo {author} {\bibfnamefont {S.}~\bibnamefont {Malick}},
  \bibinfo {author} {\bibfnamefont {R.}~\bibnamefont {Singha}}, \bibinfo
  {author} {\bibfnamefont {P.}~\bibnamefont {Mandal}}, \bibinfo {author}
  {\bibfnamefont {P.}~\bibnamefont {Rambabu}}, \bibinfo {author} {\bibfnamefont
  {V.}~\bibnamefont {Kanchana}}, \ and\ \bibinfo {author} {\bibfnamefont
  {Z.}~\bibnamefont {Hossain}},\ }\href {\doibase 10.1103/PhysRevB.99.241102}
  {\bibfield  {journal} {\bibinfo  {journal} {Phys. Rev. B}\ }\textbf {\bibinfo
  {volume} {99}},\ \bibinfo {pages} {241102} (\bibinfo {year}
  {2019})}\BibitemShut {NoStop}%
\bibitem [{\citenamefont {Laha}\ \emph
  {et~al.}(2020{\natexlab{b}})\citenamefont {Laha}, \citenamefont {Rambabu},
  \citenamefont {Kanchana}, \citenamefont {Petit}, \citenamefont {Szotek},\
  and\ \citenamefont {Hossain}}]{YbCdSn}%
  \BibitemOpen
  \bibfield  {author} {\bibinfo {author} {\bibfnamefont {A.}~\bibnamefont
  {Laha}}, \bibinfo {author} {\bibfnamefont {P.}~\bibnamefont {Rambabu}},
  \bibinfo {author} {\bibfnamefont {V.}~\bibnamefont {Kanchana}}, \bibinfo
  {author} {\bibfnamefont {L.}~\bibnamefont {Petit}}, \bibinfo {author}
  {\bibfnamefont {Z.}~\bibnamefont {Szotek}}, \ and\ \bibinfo {author}
  {\bibfnamefont {Z.}~\bibnamefont {Hossain}},\ }\href {\doibase
  10.1103/PhysRevB.102.235135} {\bibfield  {journal} {\bibinfo  {journal}
  {Phys. Rev. B}\ }\textbf {\bibinfo {volume} {102}},\ \bibinfo {pages}
  {235135} (\bibinfo {year} {2020}{\natexlab{b}})}\BibitemShut {NoStop}%
\bibitem [{\citenamefont {Kresse}\ and\ \citenamefont
  {Hafner}(1993)}]{Kresse_B47}%
  \BibitemOpen
  \bibfield  {author} {\bibinfo {author} {\bibfnamefont {G.}~\bibnamefont
  {Kresse}}\ and\ \bibinfo {author} {\bibfnamefont {J.}~\bibnamefont
  {Hafner}},\ }\href {\doibase 10.1103/PhysRevB.47.558} {\bibfield  {journal}
  {\bibinfo  {journal} {Phys. Rev. B}\ }\textbf {\bibinfo {volume} {47}},\
  \bibinfo {pages} {558} (\bibinfo {year} {1993})}\BibitemShut {NoStop}%
\bibitem [{\citenamefont {Kresse}\ and\ \citenamefont
  {Furthm\"uller}(1996{\natexlab{a}})}]{Kresse_6}%
  \BibitemOpen
  \bibfield  {author} {\bibinfo {author} {\bibfnamefont {G.}~\bibnamefont
  {Kresse}}\ and\ \bibinfo {author} {\bibfnamefont {J.}~\bibnamefont
  {Furthm\"uller}},\ }\href {\doibase 10.1016/0927-0256(96)00008-0} {\bibfield
  {journal} {\bibinfo  {journal} {Comput. Mater. Sci.}\ }\textbf {\bibinfo
  {volume} {6}},\ \bibinfo {pages} {15 } (\bibinfo {year}
  {1996}{\natexlab{a}})}\BibitemShut {NoStop}%
\bibitem [{\citenamefont {Kresse}\ and\ \citenamefont
  {Furthm\"uller}(1996{\natexlab{b}})}]{Kresse_B54}%
  \BibitemOpen
  \bibfield  {author} {\bibinfo {author} {\bibfnamefont {G.}~\bibnamefont
  {Kresse}}\ and\ \bibinfo {author} {\bibfnamefont {J.}~\bibnamefont
  {Furthm\"uller}},\ }\href {\doibase 10.1103/PhysRevB.54.11169} {\bibfield
  {journal} {\bibinfo  {journal} {Phys. Rev. B}\ }\textbf {\bibinfo {volume}
  {54}},\ \bibinfo {pages} {11169} (\bibinfo {year}
  {1996}{\natexlab{b}})}\BibitemShut {NoStop}%
\bibitem [{\citenamefont {Kresse}\ and\ \citenamefont
  {Joubert}(1999)}]{Kresse_B59}%
  \BibitemOpen
  \bibfield  {author} {\bibinfo {author} {\bibfnamefont {G.}~\bibnamefont
  {Kresse}}\ and\ \bibinfo {author} {\bibfnamefont {D.}~\bibnamefont
  {Joubert}},\ }\href {\doibase 10.1103/PhysRevB.59.1758} {\bibfield  {journal}
  {\bibinfo  {journal} {Phys. Rev. B}\ }\textbf {\bibinfo {volume} {59}},\
  \bibinfo {pages} {1758} (\bibinfo {year} {1999})}\BibitemShut {NoStop}%
\bibitem [{\citenamefont {Perdew}\ \emph {et~al.}(1996)\citenamefont {Perdew},
  \citenamefont {Burke},\ and\ \citenamefont {Ernzerhof}}]{Perdew}%
  \BibitemOpen
  \bibfield  {author} {\bibinfo {author} {\bibfnamefont {J.~P.}\ \bibnamefont
  {Perdew}}, \bibinfo {author} {\bibfnamefont {K.}~\bibnamefont {Burke}}, \
  and\ \bibinfo {author} {\bibfnamefont {M.}~\bibnamefont {Ernzerhof}},\ }\href
  {\doibase 10.1103/PhysRevLett.77.3865} {\bibfield  {journal} {\bibinfo
  {journal} {Phys. Rev. Lett.}\ }\textbf {\bibinfo {volume} {77}},\ \bibinfo
  {pages} {3865} (\bibinfo {year} {1996})}\BibitemShut {NoStop}%
\bibitem [{\citenamefont {Wan}\ \emph {et~al.}(2011)\citenamefont {Wan},
  \citenamefont {Dong},\ and\ \citenamefont {Savrasov}}]{WanU7}%
  \BibitemOpen
  \bibfield  {author} {\bibinfo {author} {\bibfnamefont {X.}~\bibnamefont
  {Wan}}, \bibinfo {author} {\bibfnamefont {J.}~\bibnamefont {Dong}}, \ and\
  \bibinfo {author} {\bibfnamefont {S.~Y.}\ \bibnamefont {Savrasov}},\ }\href
  {\doibase 10.1103/PhysRevB.83.205201} {\bibfield  {journal} {\bibinfo
  {journal} {Phys. Rev. B}\ }\textbf {\bibinfo {volume} {83}},\ \bibinfo
  {pages} {205201} (\bibinfo {year} {2011})}\BibitemShut {NoStop}%
\bibitem [{\citenamefont {Kune\ifmmode~\check{s}\else \v{s}\fi{}}\ and\
  \citenamefont {Laskowski}(2004)}]{KuneU7}%
  \BibitemOpen
  \bibfield  {author} {\bibinfo {author} {\bibfnamefont {J.}~\bibnamefont
  {Kune\ifmmode~\check{s}\else \v{s}\fi{}}}\ and\ \bibinfo {author}
  {\bibfnamefont {R.}~\bibnamefont {Laskowski}},\ }\href {\doibase
  10.1103/PhysRevB.70.174415} {\bibfield  {journal} {\bibinfo  {journal} {Phys.
  Rev. B}\ }\textbf {\bibinfo {volume} {70}},\ \bibinfo {pages} {174415}
  (\bibinfo {year} {2004})}\BibitemShut {NoStop}%
\bibitem [{\citenamefont {Monkhorst}\ and\ \citenamefont
  {Pack}(1976)}]{Monkhorst}%
  \BibitemOpen
  \bibfield  {author} {\bibinfo {author} {\bibfnamefont {H.~J.}\ \bibnamefont
  {Monkhorst}}\ and\ \bibinfo {author} {\bibfnamefont {J.~D.}\ \bibnamefont
  {Pack}},\ }\href {\doibase 10.1103/PhysRevB.13.5188} {\bibfield  {journal}
  {\bibinfo  {journal} {Phys. Rev. B}\ }\textbf {\bibinfo {volume} {13}},\
  \bibinfo {pages} {5188} (\bibinfo {year} {1976})}\BibitemShut {NoStop}%
\bibitem [{\citenamefont {Pizzi}\ \emph {et~al.}(2020)\citenamefont {Pizzi},
  \citenamefont {Vitale}, \citenamefont {Arita}, \citenamefont {Blügel},
  \citenamefont {Freimuth}, \citenamefont {G{\'{e}}ranton}, \citenamefont
  {Gibertini}, \citenamefont {Gresch}, \citenamefont {Johnson}, \citenamefont
  {Koretsune}, \citenamefont {Iba{\~{n}}ez-Azpiroz}, \citenamefont {Lee},
  \citenamefont {Lihm}, \citenamefont {Marchand}, \citenamefont {Marrazzo},
  \citenamefont {Mokrousov}, \citenamefont {Mustafa}, \citenamefont {Nohara},
  \citenamefont {Nomura}, \citenamefont {Paulatto}, \citenamefont
  {Ponc{\'{e}}}, \citenamefont {Ponweiser}, \citenamefont {Qiao}, \citenamefont
  {Thöle}, \citenamefont {Tsirkin}, \citenamefont {Wierzbowska}, \citenamefont
  {Marzari}, \citenamefont {Vanderbilt}, \citenamefont {Souza}, \citenamefont
  {Mostofi},\ and\ \citenamefont {Yates}}]{G_Pizzi}%
  \BibitemOpen
  \bibfield  {author} {\bibinfo {author} {\bibfnamefont {G.}~\bibnamefont
  {Pizzi}}, \bibinfo {author} {\bibfnamefont {V.}~\bibnamefont {Vitale}},
  \bibinfo {author} {\bibfnamefont {R.}~\bibnamefont {Arita}}, \bibinfo
  {author} {\bibfnamefont {S.}~\bibnamefont {Blügel}}, \bibinfo {author}
  {\bibfnamefont {F.}~\bibnamefont {Freimuth}}, \bibinfo {author}
  {\bibfnamefont {G.}~\bibnamefont {G{\'{e}}ranton}}, \bibinfo {author}
  {\bibfnamefont {M.}~\bibnamefont {Gibertini}}, \bibinfo {author}
  {\bibfnamefont {D.}~\bibnamefont {Gresch}}, \bibinfo {author} {\bibfnamefont
  {C.}~\bibnamefont {Johnson}}, \bibinfo {author} {\bibfnamefont
  {T.}~\bibnamefont {Koretsune}}, \bibinfo {author} {\bibfnamefont
  {J.}~\bibnamefont {Iba{\~{n}}ez-Azpiroz}}, \bibinfo {author} {\bibfnamefont
  {H.}~\bibnamefont {Lee}}, \bibinfo {author} {\bibfnamefont {J.-M.}\
  \bibnamefont {Lihm}}, \bibinfo {author} {\bibfnamefont {D.}~\bibnamefont
  {Marchand}}, \bibinfo {author} {\bibfnamefont {A.}~\bibnamefont {Marrazzo}},
  \bibinfo {author} {\bibfnamefont {Y.}~\bibnamefont {Mokrousov}}, \bibinfo
  {author} {\bibfnamefont {J.~I.}\ \bibnamefont {Mustafa}}, \bibinfo {author}
  {\bibfnamefont {Y.}~\bibnamefont {Nohara}}, \bibinfo {author} {\bibfnamefont
  {Y.}~\bibnamefont {Nomura}}, \bibinfo {author} {\bibfnamefont
  {L.}~\bibnamefont {Paulatto}}, \bibinfo {author} {\bibfnamefont
  {S.}~\bibnamefont {Ponc{\'{e}}}}, \bibinfo {author} {\bibfnamefont
  {T.}~\bibnamefont {Ponweiser}}, \bibinfo {author} {\bibfnamefont
  {J.}~\bibnamefont {Qiao}}, \bibinfo {author} {\bibfnamefont {F.}~\bibnamefont
  {Thöle}}, \bibinfo {author} {\bibfnamefont {S.~S.}\ \bibnamefont {Tsirkin}},
  \bibinfo {author} {\bibfnamefont {M.}~\bibnamefont {Wierzbowska}}, \bibinfo
  {author} {\bibfnamefont {N.}~\bibnamefont {Marzari}}, \bibinfo {author}
  {\bibfnamefont {D.}~\bibnamefont {Vanderbilt}}, \bibinfo {author}
  {\bibfnamefont {I.}~\bibnamefont {Souza}}, \bibinfo {author} {\bibfnamefont
  {A.~A.}\ \bibnamefont {Mostofi}}, \ and\ \bibinfo {author} {\bibfnamefont
  {J.~R.}\ \bibnamefont {Yates}},\ }\href {\doibase 10.1088/1361-648x/ab51ff}
  {\bibfield  {journal} {\bibinfo  {journal} {Journal of Physics: Condensed
  Matter}\ }\textbf {\bibinfo {volume} {32}},\ \bibinfo {pages} {165902}
  (\bibinfo {year} {2020})}\BibitemShut {NoStop}%
\bibitem [{\citenamefont {Wu}\ \emph {et~al.}(2018)\citenamefont {Wu},
  \citenamefont {Zhang}, \citenamefont {Song}, \citenamefont {Troyer},\ and\
  \citenamefont {Soluyanov}}]{Q_Wu}%
  \BibitemOpen
  \bibfield  {author} {\bibinfo {author} {\bibfnamefont {Q.}~\bibnamefont
  {Wu}}, \bibinfo {author} {\bibfnamefont {S.}~\bibnamefont {Zhang}}, \bibinfo
  {author} {\bibfnamefont {H.-F.}\ \bibnamefont {Song}}, \bibinfo {author}
  {\bibfnamefont {M.}~\bibnamefont {Troyer}}, \ and\ \bibinfo {author}
  {\bibfnamefont {A.~A.}\ \bibnamefont {Soluyanov}},\ }\href {\doibase
  10.1016/j.cpc.2017.09.033} {\bibfield  {journal} {\bibinfo  {journal}
  {Computer Physics Communications}\ }\textbf {\bibinfo {volume} {224}},\
  \bibinfo {pages} {405–416} (\bibinfo {year} {2018})}\BibitemShut {NoStop}%
\bibitem [{\citenamefont {Sancho}\ \emph {et~al.}(1985)\citenamefont {Sancho},
  \citenamefont {Sancho}, \citenamefont {Sancho},\ and\ \citenamefont
  {Rubio}}]{MPL}%
  \BibitemOpen
  \bibfield  {author} {\bibinfo {author} {\bibfnamefont {M.~P.~L.}\
  \bibnamefont {Sancho}}, \bibinfo {author} {\bibfnamefont {J.~M.~L.}\
  \bibnamefont {Sancho}}, \bibinfo {author} {\bibfnamefont {J.~M.~L.}\
  \bibnamefont {Sancho}}, \ and\ \bibinfo {author} {\bibfnamefont
  {J.}~\bibnamefont {Rubio}},\ }\href {\doibase 10.1088/0305-4608/15/4/009}
  {\bibfield  {journal} {\bibinfo  {journal} {Journal of Physics F: Metal
  Physics}\ }\textbf {\bibinfo {volume} {15}},\ \bibinfo {pages} {851}
  (\bibinfo {year} {1985})}\BibitemShut {NoStop}%
\bibitem [{\citenamefont {Pakhira}\ \emph {et~al.}(2020)\citenamefont
  {Pakhira}, \citenamefont {Tanatar},\ and\ \citenamefont
  {Johnston}}]{EuMg2Bi2}%
  \BibitemOpen
  \bibfield  {author} {\bibinfo {author} {\bibfnamefont {S.}~\bibnamefont
  {Pakhira}}, \bibinfo {author} {\bibfnamefont {M.~A.}\ \bibnamefont
  {Tanatar}}, \ and\ \bibinfo {author} {\bibfnamefont {D.~C.}\ \bibnamefont
  {Johnston}},\ }\href {\doibase 10.1103/PhysRevB.101.214407} {\bibfield
  {journal} {\bibinfo  {journal} {Phys. Rev. B}\ }\textbf {\bibinfo {volume}
  {101}},\ \bibinfo {pages} {214407} (\bibinfo {year} {2020})}\BibitemShut
  {NoStop}%
\bibitem [{\citenamefont {Islam}\ \emph {et~al.}(1999)\citenamefont {Islam},
  \citenamefont {Detlefs}, \citenamefont {Song}, \citenamefont {Goldman},
  \citenamefont {Antropov}, \citenamefont {Harmon}, \citenamefont {Bud'ko},
  \citenamefont {Wiener}, \citenamefont {Canfield}, \citenamefont {Wermeille},\
  and\ \citenamefont {Finkelstein}}]{RNi2Ge2}%
  \BibitemOpen
  \bibfield  {author} {\bibinfo {author} {\bibfnamefont {Z.}~\bibnamefont
  {Islam}}, \bibinfo {author} {\bibfnamefont {C.}~\bibnamefont {Detlefs}},
  \bibinfo {author} {\bibfnamefont {C.}~\bibnamefont {Song}}, \bibinfo {author}
  {\bibfnamefont {A.~I.}\ \bibnamefont {Goldman}}, \bibinfo {author}
  {\bibfnamefont {V.}~\bibnamefont {Antropov}}, \bibinfo {author}
  {\bibfnamefont {B.~N.}\ \bibnamefont {Harmon}}, \bibinfo {author}
  {\bibfnamefont {S.~L.}\ \bibnamefont {Bud'ko}}, \bibinfo {author}
  {\bibfnamefont {T.}~\bibnamefont {Wiener}}, \bibinfo {author} {\bibfnamefont
  {P.~C.}\ \bibnamefont {Canfield}}, \bibinfo {author} {\bibfnamefont
  {D.}~\bibnamefont {Wermeille}}, \ and\ \bibinfo {author} {\bibfnamefont
  {K.~D.}\ \bibnamefont {Finkelstein}},\ }\href {\doibase
  10.1103/PhysRevLett.83.2817} {\bibfield  {journal} {\bibinfo  {journal}
  {Phys. Rev. Lett.}\ }\textbf {\bibinfo {volume} {83}},\ \bibinfo {pages}
  {2817} (\bibinfo {year} {1999})}\BibitemShut {NoStop}%
\bibitem [{\citenamefont {Paramanik}\ \emph
  {et~al.}(2014{\natexlab{b}})\citenamefont {Paramanik}, \citenamefont
  {Prasad}, \citenamefont {Geibel},\ and\ \citenamefont {Hossain}}]{EuCr2As2}%
  \BibitemOpen
  \bibfield  {author} {\bibinfo {author} {\bibfnamefont {U.~B.}\ \bibnamefont
  {Paramanik}}, \bibinfo {author} {\bibfnamefont {R.}~\bibnamefont {Prasad}},
  \bibinfo {author} {\bibfnamefont {C.}~\bibnamefont {Geibel}}, \ and\ \bibinfo
  {author} {\bibfnamefont {Z.}~\bibnamefont {Hossain}},\ }\href {\doibase
  10.1103/PhysRevB.89.144423} {\bibfield  {journal} {\bibinfo  {journal} {Phys.
  Rev. B}\ }\textbf {\bibinfo {volume} {89}},\ \bibinfo {pages} {144423}
  (\bibinfo {year} {2014}{\natexlab{b}})}\BibitemShut {NoStop}%
\bibitem [{\citenamefont {Marshall}\ \emph {et~al.}(2021)\citenamefont
  {Marshall}, \citenamefont {Pletikosić}, \citenamefont {Yahyavi},
  \citenamefont {Tien}, \citenamefont {Chang}, \citenamefont {Cao},\ and\
  \citenamefont {Xie}}]{EuMg2Bi2_JAP_2021}%
  \BibitemOpen
  \bibfield  {author} {\bibinfo {author} {\bibfnamefont {M.}~\bibnamefont
  {Marshall}}, \bibinfo {author} {\bibfnamefont {I.}~\bibnamefont
  {Pletikosić}}, \bibinfo {author} {\bibfnamefont {M.}~\bibnamefont
  {Yahyavi}}, \bibinfo {author} {\bibfnamefont {H.-J.}\ \bibnamefont {Tien}},
  \bibinfo {author} {\bibfnamefont {T.-R.}\ \bibnamefont {Chang}}, \bibinfo
  {author} {\bibfnamefont {H.}~\bibnamefont {Cao}}, \ and\ \bibinfo {author}
  {\bibfnamefont {W.}~\bibnamefont {Xie}},\ }\href {\doibase 10.1063/5.0035703}
  {\bibfield  {journal} {\bibinfo  {journal} {Journal of Applied Physics}\
  }\textbf {\bibinfo {volume} {129}},\ \bibinfo {pages} {035106} (\bibinfo
  {year} {2021})}\BibitemShut {NoStop}%
\bibitem [{\citenamefont {Ma}\ \emph {et~al.}(2019)\citenamefont {Ma},
  \citenamefont {Nie}, \citenamefont {Yi}, \citenamefont {Jandke},
  \citenamefont {Shang}, \citenamefont {Yao}, \citenamefont {Naamneh},
  \citenamefont {Yan}, \citenamefont {Sun}, \citenamefont {Chikina},
  \citenamefont {Strocov}, \citenamefont {Medarde}, \citenamefont {Song},
  \citenamefont {Xiong}, \citenamefont {Xu}, \citenamefont {Wulfhekel},
  \citenamefont {Mesot}, \citenamefont {Reticcioli}, \citenamefont {Franchini},
  \citenamefont {Mudry}, \citenamefont {M{\"u}ller}, \citenamefont {Shi},
  \citenamefont {Qian}, \citenamefont {Ding},\ and\ \citenamefont
  {Shi}}]{EuCd2As2}%
  \BibitemOpen
  \bibfield  {author} {\bibinfo {author} {\bibfnamefont {J.-Z.}\ \bibnamefont
  {Ma}}, \bibinfo {author} {\bibfnamefont {S.~M.}\ \bibnamefont {Nie}},
  \bibinfo {author} {\bibfnamefont {C.~J.}\ \bibnamefont {Yi}}, \bibinfo
  {author} {\bibfnamefont {J.}~\bibnamefont {Jandke}}, \bibinfo {author}
  {\bibfnamefont {T.}~\bibnamefont {Shang}}, \bibinfo {author} {\bibfnamefont
  {M.~Y.}\ \bibnamefont {Yao}}, \bibinfo {author} {\bibfnamefont
  {M.}~\bibnamefont {Naamneh}}, \bibinfo {author} {\bibfnamefont {L.~Q.}\
  \bibnamefont {Yan}}, \bibinfo {author} {\bibfnamefont {Y.}~\bibnamefont
  {Sun}}, \bibinfo {author} {\bibfnamefont {A.}~\bibnamefont {Chikina}},
  \bibinfo {author} {\bibfnamefont {V.~N.}\ \bibnamefont {Strocov}}, \bibinfo
  {author} {\bibfnamefont {M.}~\bibnamefont {Medarde}}, \bibinfo {author}
  {\bibfnamefont {M.}~\bibnamefont {Song}}, \bibinfo {author} {\bibfnamefont
  {Y.-M.}\ \bibnamefont {Xiong}}, \bibinfo {author} {\bibfnamefont
  {G.}~\bibnamefont {Xu}}, \bibinfo {author} {\bibfnamefont {W.}~\bibnamefont
  {Wulfhekel}}, \bibinfo {author} {\bibfnamefont {J.}~\bibnamefont {Mesot}},
  \bibinfo {author} {\bibfnamefont {M.}~\bibnamefont {Reticcioli}}, \bibinfo
  {author} {\bibfnamefont {C.}~\bibnamefont {Franchini}}, \bibinfo {author}
  {\bibfnamefont {C.}~\bibnamefont {Mudry}}, \bibinfo {author} {\bibfnamefont
  {M.}~\bibnamefont {M{\"u}ller}}, \bibinfo {author} {\bibfnamefont {Y.~G.}\
  \bibnamefont {Shi}}, \bibinfo {author} {\bibfnamefont {T.}~\bibnamefont
  {Qian}}, \bibinfo {author} {\bibfnamefont {H.}~\bibnamefont {Ding}}, \ and\
  \bibinfo {author} {\bibfnamefont {M.}~\bibnamefont {Shi}},\ }\href {\doibase
  10.1126/sciadv.aaw4718} {\bibfield  {journal} {\bibinfo  {journal} {Science
  Advances}\ }\textbf {\bibinfo {volume} {5}} (\bibinfo {year} {2019}),\
  10.1126/sciadv.aaw4718}\BibitemShut {NoStop}%
\bibitem [{\citenamefont {Xu}\ \emph {et~al.}(2020{\natexlab{a}})\citenamefont
  {Xu}, \citenamefont {Wang}, \citenamefont {Wang}, \citenamefont {Su},
  \citenamefont {Wang},\ and\ \citenamefont {Xia}}]{BaAgAs_family}%
  \BibitemOpen
  \bibfield  {author} {\bibinfo {author} {\bibfnamefont {S.}~\bibnamefont
  {Xu}}, \bibinfo {author} {\bibfnamefont {H.}~\bibnamefont {Wang}}, \bibinfo
  {author} {\bibfnamefont {Y.-Y.}\ \bibnamefont {Wang}}, \bibinfo {author}
  {\bibfnamefont {Y.}~\bibnamefont {Su}}, \bibinfo {author} {\bibfnamefont
  {X.-Y.}\ \bibnamefont {Wang}}, \ and\ \bibinfo {author} {\bibfnamefont
  {T.-L.}\ \bibnamefont {Xia}},\ }\href {\doibase
  https://doi.org/10.1016/j.jcrysgro.2019.125304} {\bibfield  {journal}
  {\bibinfo  {journal} {Journal of Crystal Growth}\ }\textbf {\bibinfo {volume}
  {531}},\ \bibinfo {pages} {125304} (\bibinfo {year}
  {2020}{\natexlab{a}})}\BibitemShut {NoStop}%
\bibitem [{\citenamefont {Ren}\ \emph {et~al.}(2010)\citenamefont {Ren},
  \citenamefont {Taskin}, \citenamefont {Sasaki}, \citenamefont {Segawa},\ and\
  \citenamefont {Ando}}]{Bi2Te2Se}%
  \BibitemOpen
  \bibfield  {author} {\bibinfo {author} {\bibfnamefont {Z.}~\bibnamefont
  {Ren}}, \bibinfo {author} {\bibfnamefont {A.~A.}\ \bibnamefont {Taskin}},
  \bibinfo {author} {\bibfnamefont {S.}~\bibnamefont {Sasaki}}, \bibinfo
  {author} {\bibfnamefont {K.}~\bibnamefont {Segawa}}, \ and\ \bibinfo {author}
  {\bibfnamefont {Y.}~\bibnamefont {Ando}},\ }\href {\doibase
  10.1103/PhysRevB.82.241306} {\bibfield  {journal} {\bibinfo  {journal} {Phys.
  Rev. B}\ }\textbf {\bibinfo {volume} {82}},\ \bibinfo {pages} {241306}
  (\bibinfo {year} {2010})}\BibitemShut {NoStop}%
\bibitem [{\citenamefont {Sato}\ \emph {et~al.}(2021)\citenamefont {Sato},
  \citenamefont {Xiang}, \citenamefont {Kasahara}, \citenamefont {Kasahara},
  \citenamefont {Chen}, \citenamefont {Tinsman}, \citenamefont {Iga},
  \citenamefont {Singleton}, \citenamefont {Nair}, \citenamefont {Maksimovic},
  \citenamefont {Analytis}, \citenamefont {Li},\ and\ \citenamefont
  {Matsuda}}]{YbB12}%
  \BibitemOpen
  \bibfield  {author} {\bibinfo {author} {\bibfnamefont {Y.}~\bibnamefont
  {Sato}}, \bibinfo {author} {\bibfnamefont {Z.}~\bibnamefont {Xiang}},
  \bibinfo {author} {\bibfnamefont {Y.}~\bibnamefont {Kasahara}}, \bibinfo
  {author} {\bibfnamefont {S.}~\bibnamefont {Kasahara}}, \bibinfo {author}
  {\bibfnamefont {L.}~\bibnamefont {Chen}}, \bibinfo {author} {\bibfnamefont
  {C.}~\bibnamefont {Tinsman}}, \bibinfo {author} {\bibfnamefont
  {F.}~\bibnamefont {Iga}}, \bibinfo {author} {\bibfnamefont {J.}~\bibnamefont
  {Singleton}}, \bibinfo {author} {\bibfnamefont {N.~L.}\ \bibnamefont {Nair}},
  \bibinfo {author} {\bibfnamefont {N.}~\bibnamefont {Maksimovic}}, \bibinfo
  {author} {\bibfnamefont {J.~G.}\ \bibnamefont {Analytis}}, \bibinfo {author}
  {\bibfnamefont {L.}~\bibnamefont {Li}}, \ and\ \bibinfo {author}
  {\bibfnamefont {Y.}~\bibnamefont {Matsuda}},\ }\href {\doibase
  10.1088/1361-6463/ac10d9} {\bibfield  {journal} {\bibinfo  {journal} {Journal
  of Physics D: Applied Physics}\ }\textbf {\bibinfo {volume} {54}},\ \bibinfo
  {pages} {404002} (\bibinfo {year} {2021})}\BibitemShut {NoStop}%
\bibitem [{\citenamefont {Wolgast}\ \emph {et~al.}(2013)\citenamefont
  {Wolgast}, \citenamefont {Kurdak}, \citenamefont {Sun}, \citenamefont
  {Allen}, \citenamefont {Kim},\ and\ \citenamefont {Fisk}}]{SmB6}%
  \BibitemOpen
  \bibfield  {author} {\bibinfo {author} {\bibfnamefont {S.}~\bibnamefont
  {Wolgast}}, \bibinfo {author} {\bibfnamefont {i.~m. c. b. u. i. e. i.~f.}\
  \bibnamefont {Kurdak}}, \bibinfo {author} {\bibfnamefont {K.}~\bibnamefont
  {Sun}}, \bibinfo {author} {\bibfnamefont {J.~W.}\ \bibnamefont {Allen}},
  \bibinfo {author} {\bibfnamefont {D.-J.}\ \bibnamefont {Kim}}, \ and\
  \bibinfo {author} {\bibfnamefont {Z.}~\bibnamefont {Fisk}},\ }\href {\doibase
  10.1103/PhysRevB.88.180405} {\bibfield  {journal} {\bibinfo  {journal} {Phys.
  Rev. B}\ }\textbf {\bibinfo {volume} {88}},\ \bibinfo {pages} {180405}
  (\bibinfo {year} {2013})}\BibitemShut {NoStop}%
\bibitem [{\citenamefont {Shon}\ \emph {et~al.}(2019)\citenamefont {Shon},
  \citenamefont {Rhyee}, \citenamefont {Jin},\ and\ \citenamefont
  {Kim}}]{EuBiTe3}%
  \BibitemOpen
  \bibfield  {author} {\bibinfo {author} {\bibfnamefont {W.}~\bibnamefont
  {Shon}}, \bibinfo {author} {\bibfnamefont {J.-S.}\ \bibnamefont {Rhyee}},
  \bibinfo {author} {\bibfnamefont {Y.}~\bibnamefont {Jin}}, \ and\ \bibinfo
  {author} {\bibfnamefont {S.-J.}\ \bibnamefont {Kim}},\ }\href {\doibase
  10.1103/PhysRevB.100.024433} {\bibfield  {journal} {\bibinfo  {journal}
  {Phys. Rev. B}\ }\textbf {\bibinfo {volume} {100}},\ \bibinfo {pages}
  {024433} (\bibinfo {year} {2019})}\BibitemShut {NoStop}%
\bibitem [{\citenamefont {Xu}\ \emph {et~al.}(2020{\natexlab{b}})\citenamefont
  {Xu}, \citenamefont {Xi},\ and\ \citenamefont {Gao}}]{Heja}%
  \BibitemOpen
  \bibfield  {author} {\bibinfo {author} {\bibfnamefont {H.}~\bibnamefont
  {Xu}}, \bibinfo {author} {\bibfnamefont {H.}~\bibnamefont {Xi}}, \ and\
  \bibinfo {author} {\bibfnamefont {Y.-C.}\ \bibnamefont {Gao}},\ }\href
  {\doibase 10.3389/fchem.2020.608398} {\bibfield  {journal} {\bibinfo
  {journal} {Frontiers in Chemistry}\ }\textbf {\bibinfo {volume} {8}},\
  \bibinfo {pages} {1003} (\bibinfo {year} {2020}{\natexlab{b}})}\BibitemShut
  {NoStop}%
\end{thebibliography}%
	
\end{document}